\documentclass{statsoc}
\usepackage[a4paper]{geometry}
\usepackage[utf8]{inputenc}
\usepackage{amsfonts,epsfig}
\usepackage[hyphens]{url}
\usepackage{hyperref}
\hypersetup{
    colorlinks=true,
    linkcolor=black,
    citecolor=black,
    filecolor=black,
    urlcolor=black,
}
\usepackage{breakurl}
\usepackage{comment}
\usepackage{color}
\usepackage{microtype}
\usepackage{subcaption}

\usepackage{relsize}
\usepackage{fancyvrb}
\usepackage{amssymb,amsmath}
\graphicspath{{plots/}}

\newcommand{\PM}{PM$_{2.5}$}
\newcommand{\bayesplot}{{\tt bayesplot}}

\title{Visualization in Bayesian workflow}

\author[Gabry {\it et al.}]{Jonah Gabry\footnote{Joint first author}}
\address{Department of Statistics and ISERP, Columbia University, New York, USA.}
\email{\small jonah.sol.gabry@columbia.edu}

\author[Gabry {\it et al.}]{Daniel Simpson\footnotemark[1]}
\address{Department of Statistical Sciences, University of Toronto, Canada.}

\author[Gabry {\it et al.}]{Aki Vehtari}
\address{Department of Computer Science, Aalto University, Espoo, Finland.}

\author[Gabry {\it et al.}]{Michael Betancourt}
\address{ISERP, Columbia University, and Symplectomorphic, LLC, New York, USA. }

\author[Gabry {\it et al.}]{Andrew Gelman}
\address{Departments of Statistics and Political Science, Columbia University, New York, USA.}

\begin{document}
\maketitle

\begin{abstract}
Bayesian data analysis is about more than just computing a posterior
distribution, and Bayesian visualization is about more than trace plots of
Markov chains. Practical Bayesian data analysis, like all data analysis, is an
iterative process of model building, inference, model checking and evaluation,
and model expansion. Visualization is helpful in each of these stages of the
Bayesian workflow and it is indispensable when drawing inferences from the types
of  modern, high-dimensional models that are used by applied researchers.
\end{abstract}

\section{Introduction and running example}
\label{sec:intro}

Visualization is a vital tool for data analysis, and its role is
well established in both the exploratory and final presentation stages of a
statistical workflow. In this paper, we argue that the same visualization tools
should be used at all points during an analysis.  We illustrate this thesis by
following a single real example, estimating the global concentration of a
certain type of air pollution, through all of the phases of statistical
workflow:
(a) Exploratory data analysis to aid in setting up an initial model;
(b) Computational model checks using fake-data simulation and the prior
      predictive distribution;
(c) Computational checks to ensure the inference algorithm works reliably,
(d) Posterior predictive checks and other juxtapositions of data and predictions
      under the fitted model;
(e)  Model comparison via tools such as cross-validation.

The tools developed in this paper are implemented in the \bayesplot\ R package
\citep{bayesplotRpackage, rcore}, which uses {\tt ggplot2} \citep{ggplot2Rpackage}
and is linked to---though not dependent on---Stan \citep{rstan,stanmanual},
the general-purpose Hamiltonian Monte Carlo engine for Bayesian model 
fitting.\footnote{Code and data for fitting the models and making the plots in this paper 
are available at \url{https://github.com/jgabry/bayes-vis-paper}}

In order to better discuss the ways visualization can aid a statistical
workflow we consider a particular problem, the estimation of human exposure to
air pollution from particulate matter measuring less than 2.5 microns in
diameter (\PM). Exposure to \PM\ is linked to a number of poor health outcomes,
and a recent report estimated that \PM\ is responsible for three million deaths
worldwide each year \citep{shaddick2017data}.

For our running example, we use the data from \citet{shaddick2017data},
aggregated to the city level, to estimate ambient \PM\ concentration across the
world. The statistical problem is that we only have direct measurements of \PM\
from a sparse network of $2980$ ground monitors with heterogeneous spatial
coverage (Figure~\ref{fig:datamap}). This monitoring network has especially poor
coverage across Africa, central Asia, and Russia.

In order to estimate the public health effect of \PM , we need estimates of the
\PM\ concentration at the same spatial resolution as the population data. To
obtain these estimates, we supplement the direct measurements with a
high-resolution satellite data product that converts measurements of aerosol
optical depth into estimates of \PM.  The hope is that we can use the ground
monitor data to calibrate the approximate satellite measurements, and hence get
estimates of \PM\ at the required spatial resolution.

The aim of this analysis is to build a predictive model of \PM\ with
appropriately calibrated prediction intervals.  We will not attempt a full
analysis of this data, which was undertaken by \citet{shaddick2017data}.
Instead, we will focus on three simple, but plausible, models for the data in
order to show how visualization can be used to help construct, sense-check,
compute, and evaluate these models.

\section{Exploratory data analysis goes beyond just plotting the data}
\label{sec:exploratory}

\begin{figure}
\begin{subfigure}{0.49\textwidth}
\includegraphics[width=\textwidth]{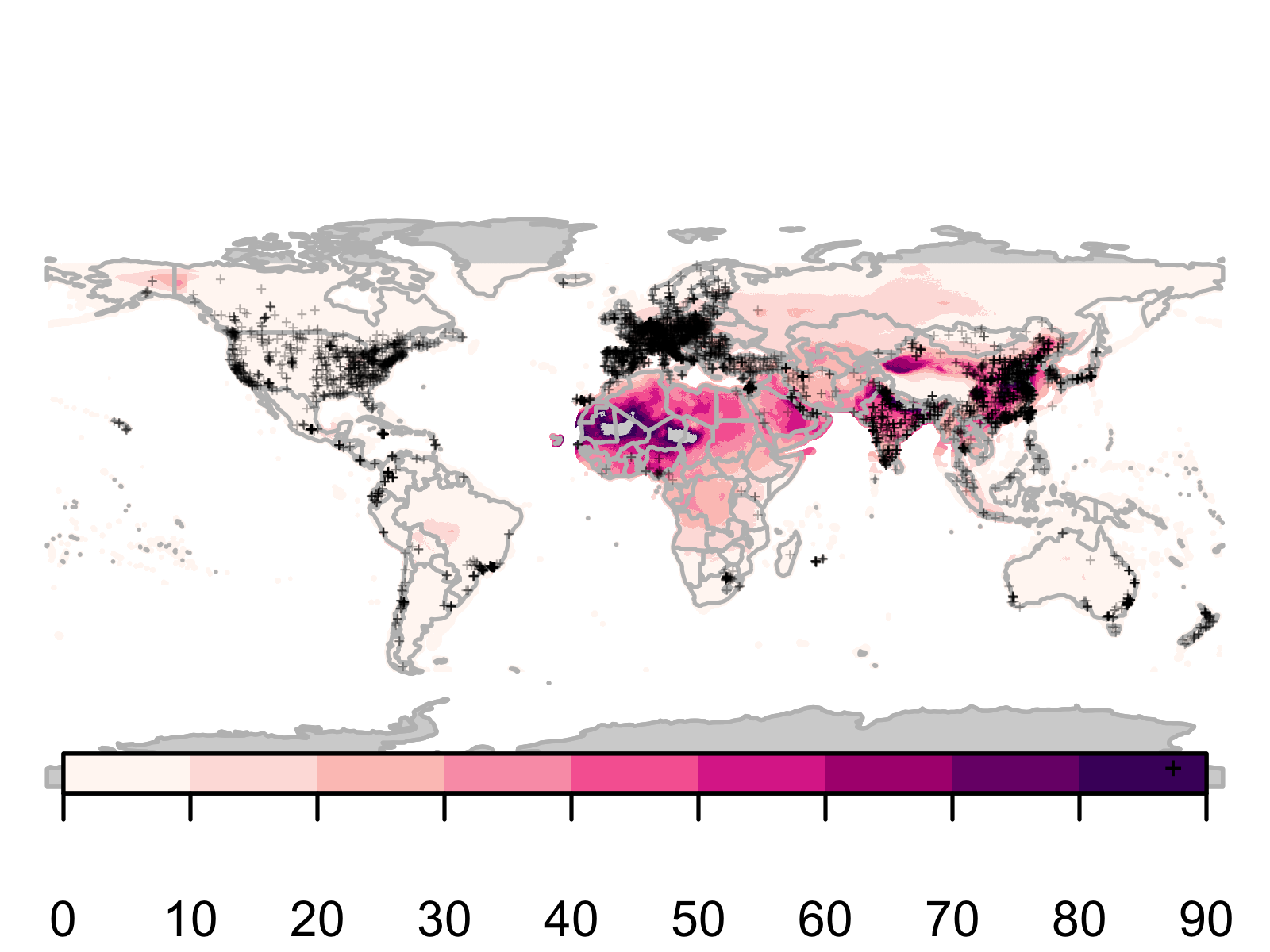}
\caption{The satellite estimates of \PM. The black points indicate the locations
the ground monitors.}
\label{fig:datamap}
\end{subfigure}
~
\begin{subfigure}{0.49\textwidth}
\includegraphics[width=\textwidth]{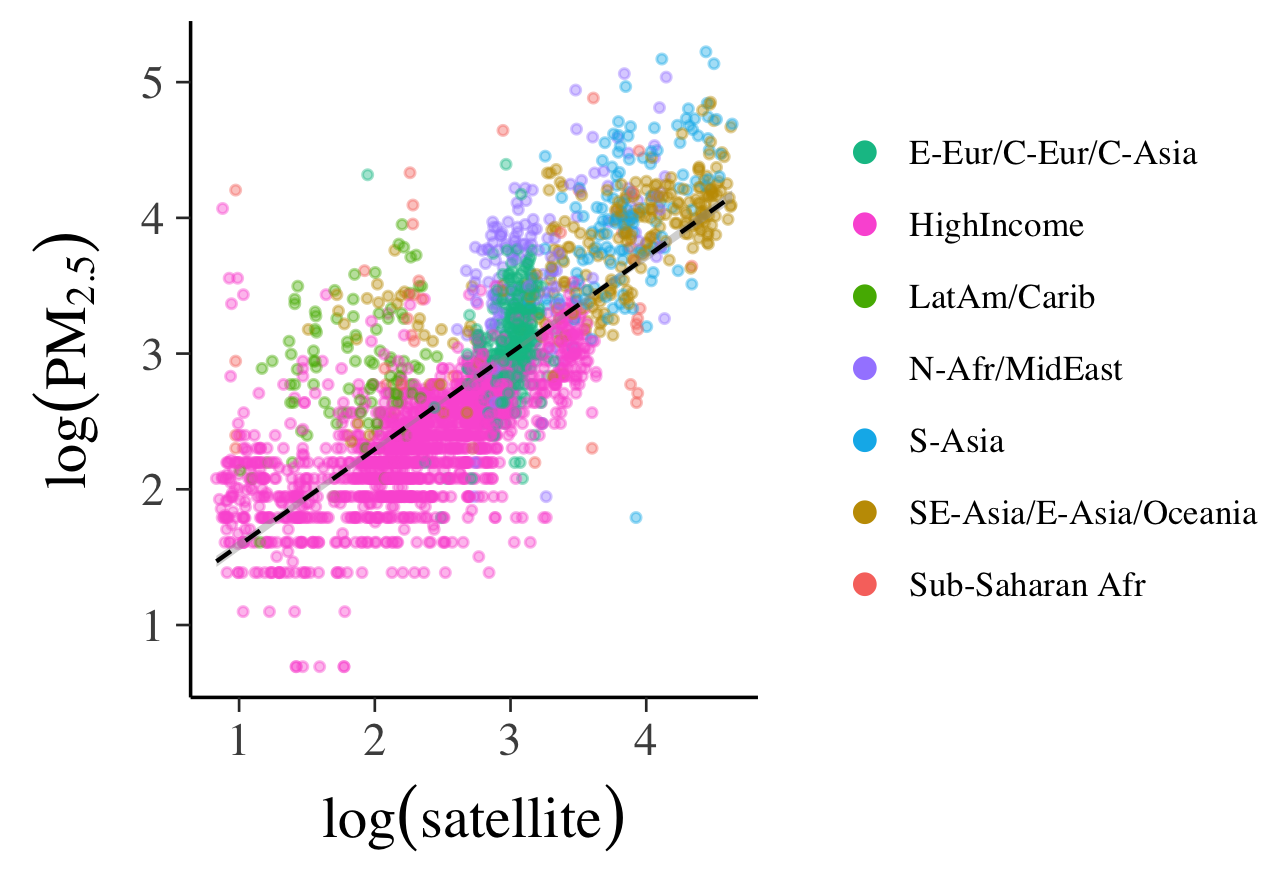}
\caption{A scatterplot of $\log($\PM$)$ vs $\log(\text{satellite})$. The points
are colored by WHO super region.
}
\label{fig:plot1}
\end{subfigure}
\caption{\it Data displays for our running example of exposure to particulate matter.}
\end{figure}

An important aspect of formalizing the role of visualization in exploratory data
analysis is to place it within the context of a particular statistical workflow.
In particular, we argue that exploratory data analysis is more than simply
plotting the data. Instead, we consider it a method to build a network of
increasingly complex models that can capture the features and heterogeneities
present in the data   \citep{gelman2004exploratory}.

This ground-up modeling strategy is particularly useful when the data that have
been gathered are sparse or unbalanced, as the resulting network of models is
built knowing the limitations of the design. A different strategy, which is
common in machine learning, is to build a top-down model that throws all
available information into a complicated non-parametric procedure.  This works
well for data that are a good representation of the population of interest but
can be prone to over-fitting or generalization error when used on sparse or
unbalanced data.  Using a purely predictive model to calibrate the satellite
measurements would yield a fit that would be dominated by data in Western Europe
and North America, which have very different air pollution profiles than most
developing nations. With this in mind, we use the ground-up strategy to build a
small network of three simple models for predicting \PM\ on a global scale.

The simplest predictive model that we can fit assumes that the satellite data
product is a good predictor of the ground monitor data after a simple affine
adjustment. In fact, this was the model used by the Global Burden of Disease
project before the 2016 update \citep{forouzanfar2015global}. Figure
\ref{fig:plot1} shows a straight line that fits the data on a log-log scale
reasonably well ($R^2 \approx 0.6$). Discretization artifacts at the lower
values of \PM\ are also clearly visible.

\begin{figure}
\centering
\begin{subfigure}{0.45\textwidth}
\includegraphics[width=\textwidth]{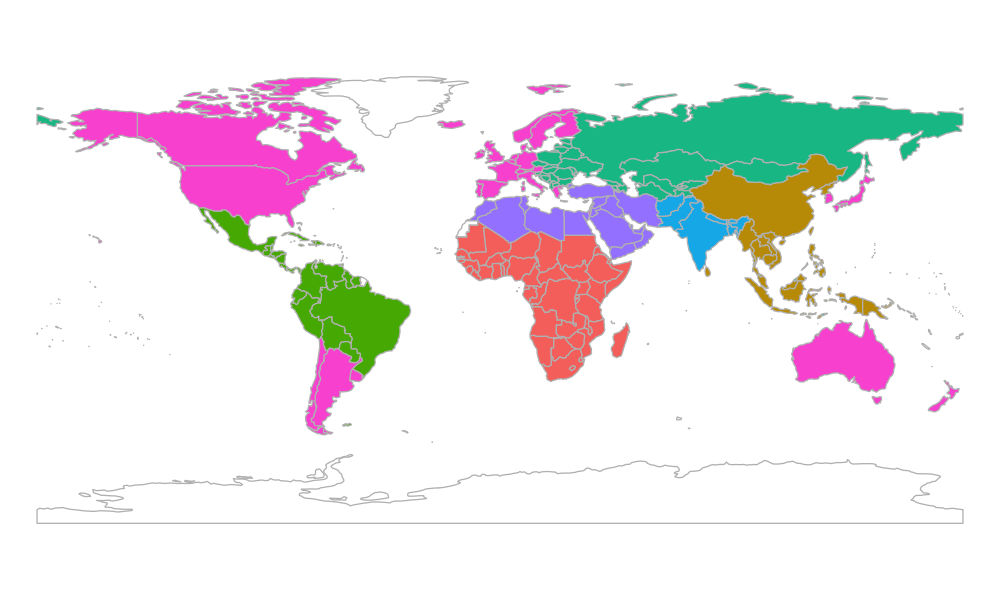}
\caption{The WHO super-regions. The pink super-region corresponds to wealthy
countries. The remaining regions are defined based on geographic contiguity.}
\label{fig:whomap}
\end{subfigure}
~
\begin{subfigure}{0.45\textwidth}
\includegraphics[width=\textwidth]{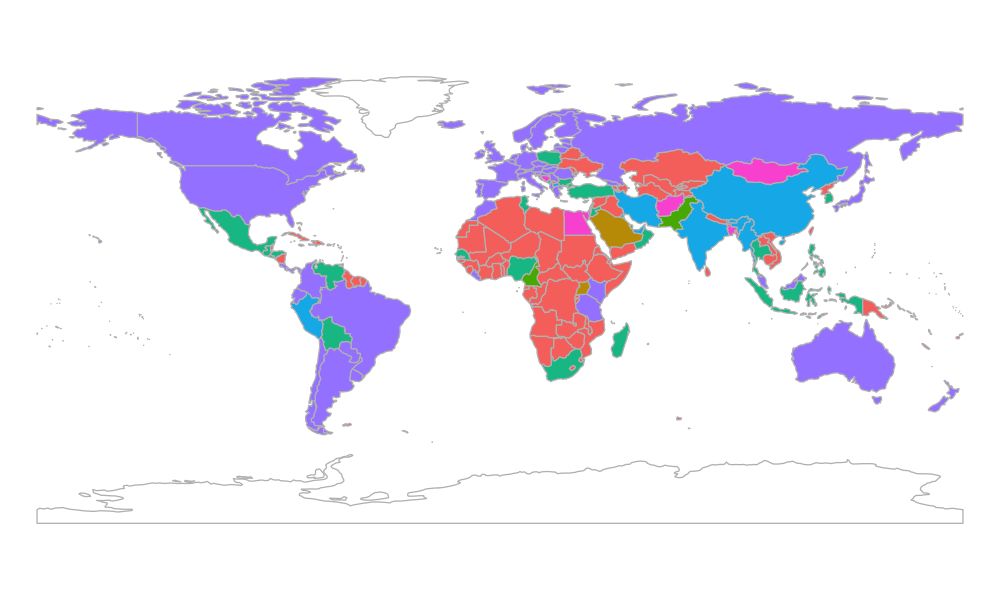}
\caption{The super-regions found by clustering based on ground measurements of
\PM. Countries for which we have no ground monitor measurements are colored
red.}
\label{fig:clustermap}
\end{subfigure}
\caption{\it World Health Organization super-regions and super-regions from
clustering.}
\end{figure}

To improve the model, we need to think about possible sources of heterogeneity.
For example, we know that developed and developing countries have different
levels of industrialization and hence different air pollution.  We also know
that desert sand can be a large source of \PM.  If these differences are not
appropriately captured by the satellite data product, fitting only a single
regression line could leave us in danger of falling prey to Simpson's paradox
(that a trend can reverse when data are grouped).

To expand out our network of models, we consider two possible groupings of
countries. The WHO super-regions (Figure~\ref{fig:whomap}) separate out rich
countries and divide the remaining countries into six geographically contiguous
regions. These regions have not been constructed with air pollution in mind, so
we also constructed a different division based on a $6$-component hierarchical
clustering of ground monitor measurements of \PM\ (Figure~\ref{fig:clustermap}).
The seventh region constructed this way is the collection of all countries for
which we do not have ground monitor data.

\begin{figure}
\centering
\begin{subfigure}{0.48\textwidth}
\includegraphics[width=\textwidth]{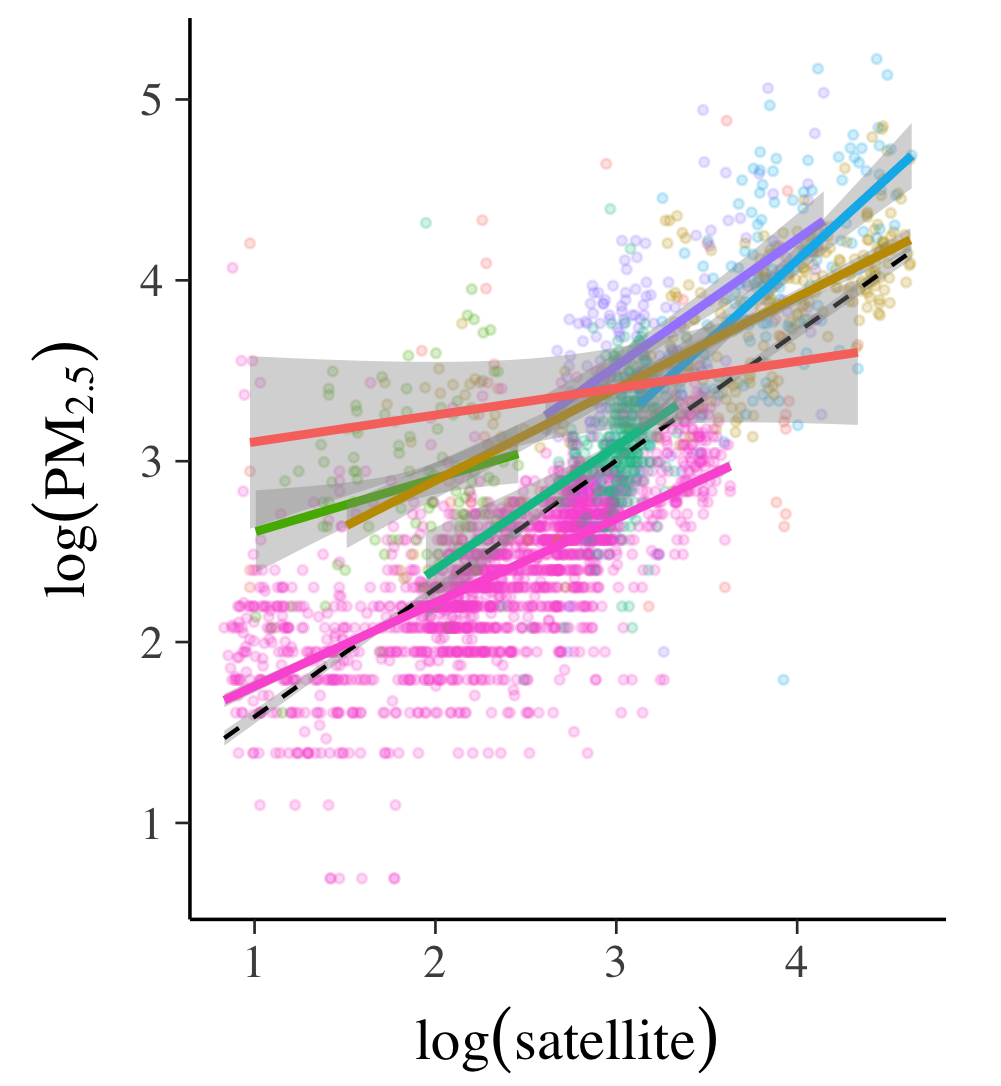}
\caption{The same as Figure \ref{fig:plot1}, but also showing independent linear
models fit within each WHO super-region.}
\label{fig:plot2}
\end{subfigure}
~
\begin{subfigure}{0.48\textwidth}
\includegraphics[width=\textwidth]{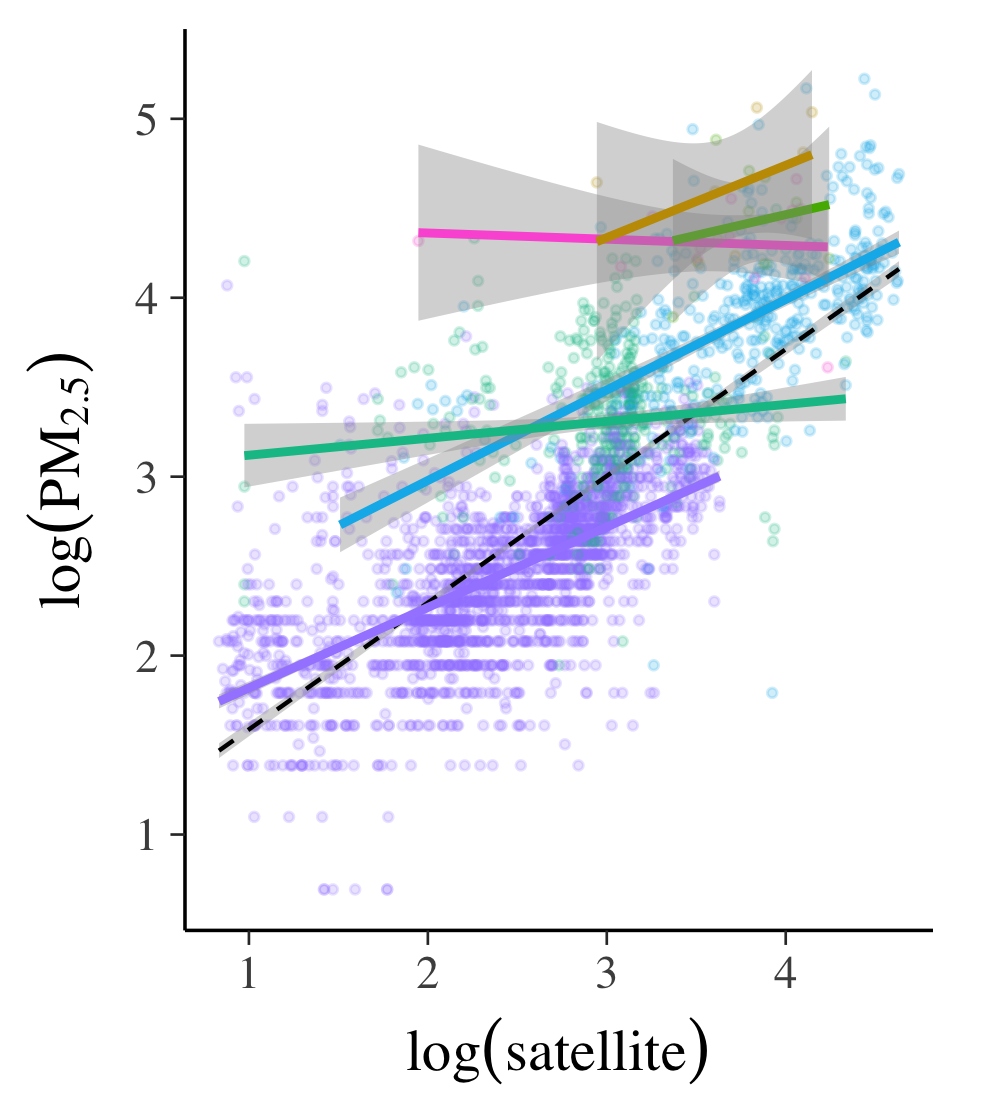}
\caption{The same as (a), but the the linear models are fit within each of the
cluster regions shown in Figure \ref{fig:clustermap}.}
\label{fig:plot3}
\end{subfigure}

\caption{\it Graphics in model building: here, evidence that a single linear
trend is insufficient.}
\end{figure}

When the trends for each of these regions are plotted individually (Figures
\ref{fig:plot2}, \ref{fig:plot3}), it is clear that some ecological bias would
enter into the analysis if we only used a single linear regression.  We also see
that some regions, particularly Sub-Saharan Africa (red in Figure~\ref{fig:plot2}) 
and clusters 1 and 6 (pink and yellow in Figure~\ref{fig:plot3}), do not have enough 
data to comprehensively pin down the linear trend. This suggests that some 
borrowing of strength through a multilevel model may be appropriate.

From this preliminary data analysis, we have constructed a network of three
potential models.  Model 1 is a simple linear regression. Model 2 is a
multilevel model where observations are stratified by WHO super-region. Model 3
is a multilevel model where observations are stratified by \emph{clustered}
super-region.

These three models will be sufficient for demonstrating our proposed workflow,
but this is a smaller network of models than we would use for a comprehensive
analysis of the \PM\ data. \citet{shaddick2017data}, for example, also consider
smaller regions, country-level variation, and a spatial model for the varying
coefficients. Further calibration covariates can also be included.

\section{Fake data can be almost as valuable as real data for building your model}
\label{sec:prior_pred}

\begin{figure}
\centering
\begin{subfigure}{0.31\textwidth}
\includegraphics[width=\textwidth]{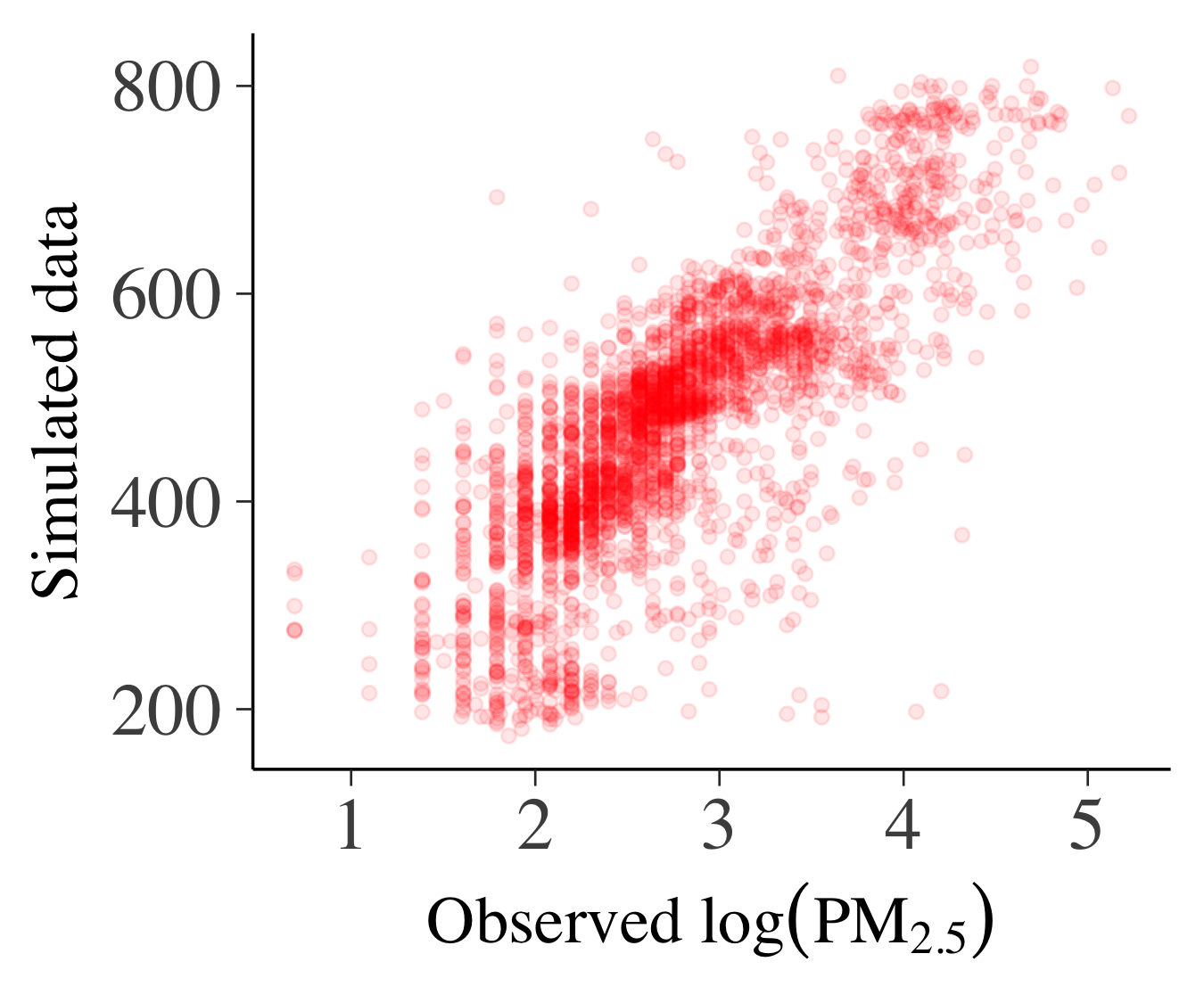}
\caption{Vague priors}
\label{fig:prior_pred_vague}
\end{subfigure}
~
\begin{subfigure}{0.31\textwidth}
\includegraphics[width=\textwidth]{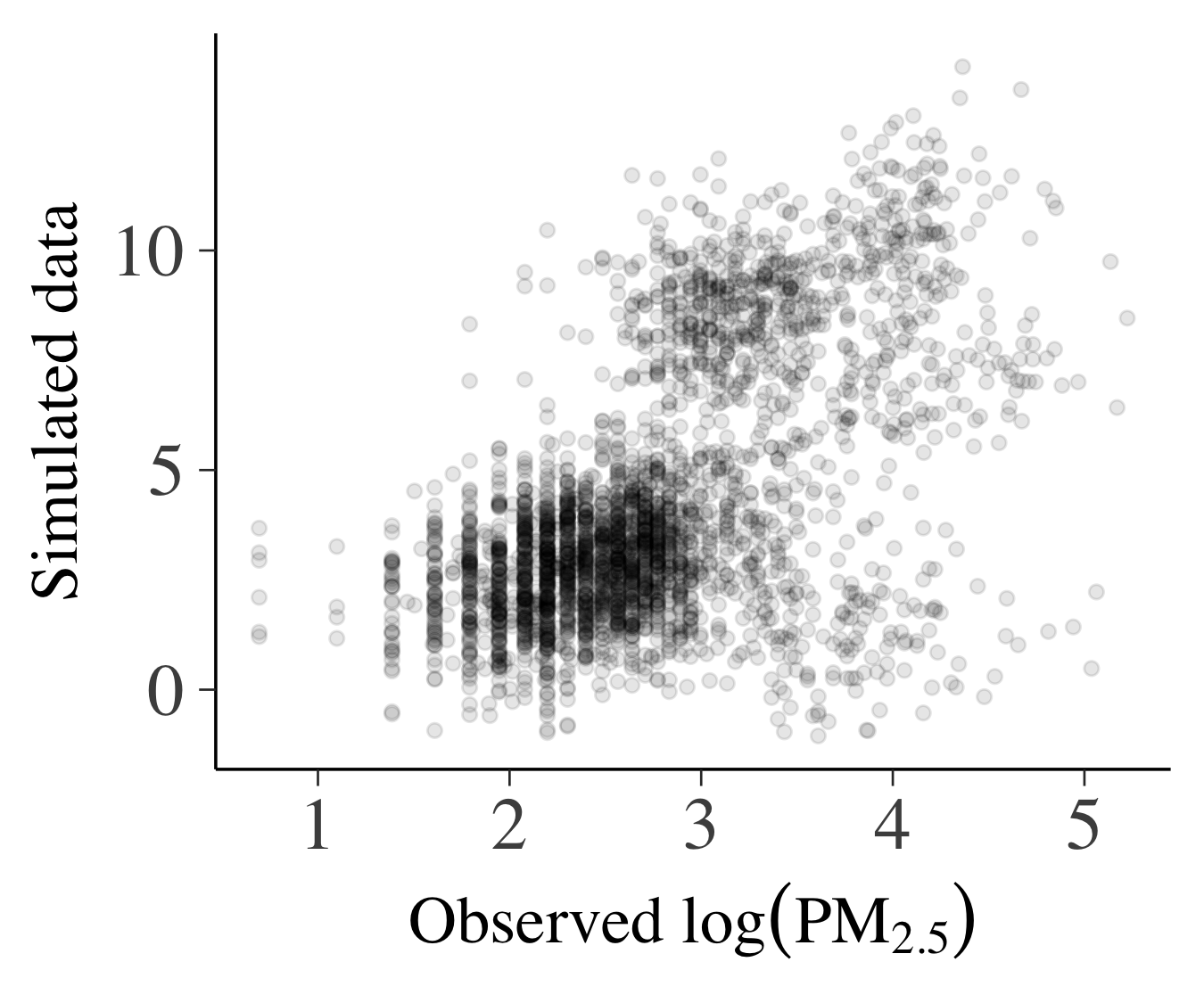}
\caption{Weakly informative priors}
\label{fig:prior_pred_wip}
\end{subfigure}
~
\begin{subfigure}{0.31\textwidth}
\includegraphics[width=\textwidth]{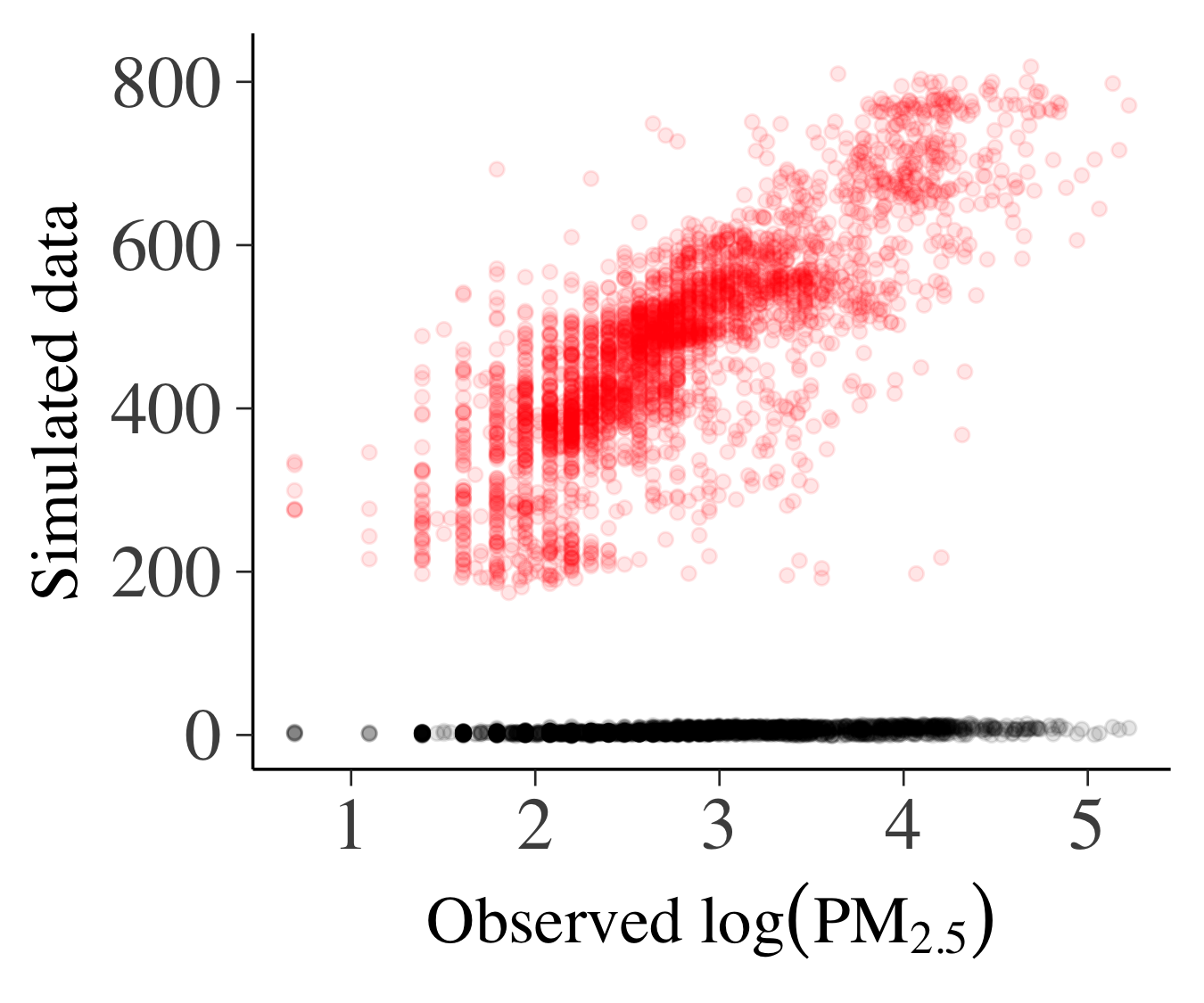}
\caption{Comparison}
\label{fig:prior_pred_compare}
\end{subfigure}

\caption{\it Visualizing the prior predictive distribution.
Panels (a) and (b) show realizations from the prior predictive distribution using
priors for the $\beta$'s and $\tau$'s that are vague and weakly informative,
respectively. The same $N_+(0,1)$ prior is used for $\sigma$ in both cases.
Simulated data are plotted on the $y$-axis and observed data on the $x$-axis.
Because the simulations under the vague and weakly informative priors are so different,
the $y$-axis scales used in panels (a) and (b) also differ dramatically.
Panel (c) emphasizes the difference in the simulations by showing the
red points from (a) and the black points from (b) plotted using the same $y$-axis.}
\label{fig:prior_pred}
\end{figure}

The exploratory data analysis resulted in a network of three models: one linear
regression model and two different linear multilevel models. In order to fully
specify these models, we need to specify prior distributions on all of the
parameters. If we specify proper priors for all parameters in the model, a
Bayesian model yields a joint prior distribution on parameters and data, and
hence a prior marginal distribution for the data.  That is, Bayesian models with
proper priors are \emph{generative models}. The main idea in this section is
that we can visualize simulations from the prior marginal distribution of the
data to assess the consistency of the chosen priors with domain knowledge.

The main advantage to assessing priors based on the prior marginal distribution
for the data is that it reflects the interplay between the prior distribution on
the parameters and the likelihood. This is a vital component of understanding
how prior distributions actually work for a given problem
\citep{gelman2017priors}. It also explicitly reflects the idea that we can't
fully understand the prior by fixing all but one parameter and assessing the
effect of the unidimensional marginal prior.  Instead, we need to assess the
effect of the prior as a multivariate distribution.

The prior distribution over the data allows us to extend the concept of a weakly
informative prior \citep{gelman2008weakly}  to be more aware of the role of the
likelihood. In particular, we say that a prior leads to a \emph{weakly
informative joint prior data generating process} if draws from the prior data
generating distribution $p(y)$ could represent any data set that could plausibly
be observed.  As with the standard concept of weakly informative priors, it is
important that this prior predictive distribution for the data has at least some
mass around extreme but plausible data sets. On the other hand, there should be
no mass on completely implausible data sets. We recommend assessing  how
informative the prior distribution on the data is by generating a ``flip book''
of simulated datasets that can be used to investigate the variability and
multivariate structure of the distribution.

To demonstrate the power of this approach, we return to the multilevel model for
the \PM\ data. Mathematically, the model will look like
$y_{ij} \sim N(\beta_0 + \beta_{0j} + (\beta_1 + \beta_{1j}) \, x_{ij}, \, \sigma^2),$
$\beta_{0j}  \sim N(0, \tau_0^2),$
$\beta_{1j} \sim N(0, \tau_1^2),$ 
where $y_{ij}$ is the logarithm of the observed \PM, $x_{ij}$ is the logarithm
of the estimate from the satellite model, $i$ ranges over the observations in
each super-region, $j$ ranges over the super-regions, and $\sigma$, $\tau_0$,
$\tau_1$, $\beta_0$ and $\beta_1$ need prior distributions.

Consider some priors of the sort that are sometimes recommended as being vague:
$\beta_k \sim N(0,100)$, $\tau_k^2 \sim \text{Inv-Gamma}(1,100)$. The data
generated using these priors and shown in Figure \ref{fig:prior_pred_vague} are
completely impossible for this application; note the $y$-axis limits and recall
that the data are on the log scale.  This is primarily because the vague priors
don't actually respect our contextual knowledge.

We know that the satellite estimates are reasonably faithful representations of
the \PM\ concentration, so a more sensible set of priors would be centered
around models with intercept 0 and slope 1.  An example of this would be
$\beta_0 \sim N(0,1)$, $\beta_1 \sim N(1,1)$, $\tau_k \sim N_+(0,1)$, 
where $N_+$ is the half-normal distribution. Data generated 
by this model are shown in Figure \ref{fig:prior_pred_wip}. While it
is clear that this realization corresponds to a mis-calibrated satellite
model (especially when we remember that we are working on the log scale), it is
quite a bit more plausible than the model with vague priors.

We argue that these tighter priors are still only {weakly} informative, in that
the implied data generating process can still generate data that is much more
extreme than we would expect from our domain knowledge.  In fact, when repeating
the simulation shown in Figure \ref{fig:prior_pred_wip} many times we found that
generating data using these priors can produce data points with more than
$22,\!000 \mu\text{g}m^{-3}$, which is a still a very high number in this
context.

The prior predictive distribution is a powerful tool for understanding the
structure of our model before we make a measurement, but its density evaluated
at the measured data also plays the role of the marginal likelihood which is
commonly used in model comparison.  Unfortunately the utility of the the prior
predictive distribution to evaluate the model does not extend to utility in
selecting between models. For further discussion see \citet{gelman2017priors}.

\section{Graphical Markov chain Monte Carlo diagnostics: moving beyond trace plots}
\label{sec:mcmc}

Constructing a network of models is only the first step in the Bayesian
workflow. Our next job is to fit them. Once again, visualizations can be a key
tool in doing this well. Traditionally, Markov chain Monte Carlo (MCMC)
diagnostic plots consist of trace plots and autocorrelation functions.  We find
these plots can be helpful to understand problems that have been caught by
numerical summaries such as the potential scale reduction factor $\widehat{R}$
\citep[Section 30.3]{stanmanual}, but they are not always needed as part of
workflow in the many settings where chains mix well.

For general MCMC methods it is difficult to do any better than between/within
summary comparisons, following up with trace plots as needed.  But if we
restrict our attention to Hamiltonian Monte Carlo (HMC) and its variants, we can
get much more detailed information about the performance of the Markov chain
\citep{betancourt2017conceptual}. We know that the success of HMC requires that
the geometry of the set containing the bulk of the posterior probability mass
(which we call the typical set)  is fairly smooth. It is not possible to check
this condition mathematically for most models, but it can be checked
numerically. It turns out that if the geometry of the typical set is non-smooth,
the  path taken by leap-frog integrator that defines the HMC proposal will
rapidly diverge from the energy conserving trajectory.

Diagnosing divergent numerical trajectories precisely is difficult, but it is
straightforward to identify these divergences heuristically by checking if the
error in the Hamiltonian crosses a large threshold. Occasionally this heuristic
falsely flags stable trajectories as divergent, but we can identify these false
positives visually by checking if the samples generated from divergent
trajectories are distributed in the same way as the non-divergent trajectories.
Combining this simple heuristic with visualization greatly increases its value.

Visually, a concentration of divergences in small neighborhoods of parameter
space, however, indicates a region of high curvature in the posterior that
obstructs exploration.  These neighborhoods will also impede any  MCMC method
based on local information, but to our knowledge only HMC has enough
mathematical structure to be able to reliably diagnose these features. Hence,
when we are using HMC for our inference, we can use visualization to
assess the convergence of the MCMC method and also to understand the 
geometry of the posterior.

\begin{figure}
\centering
\begin{subfigure}{0.48\textwidth}
\includegraphics[width=\textwidth]{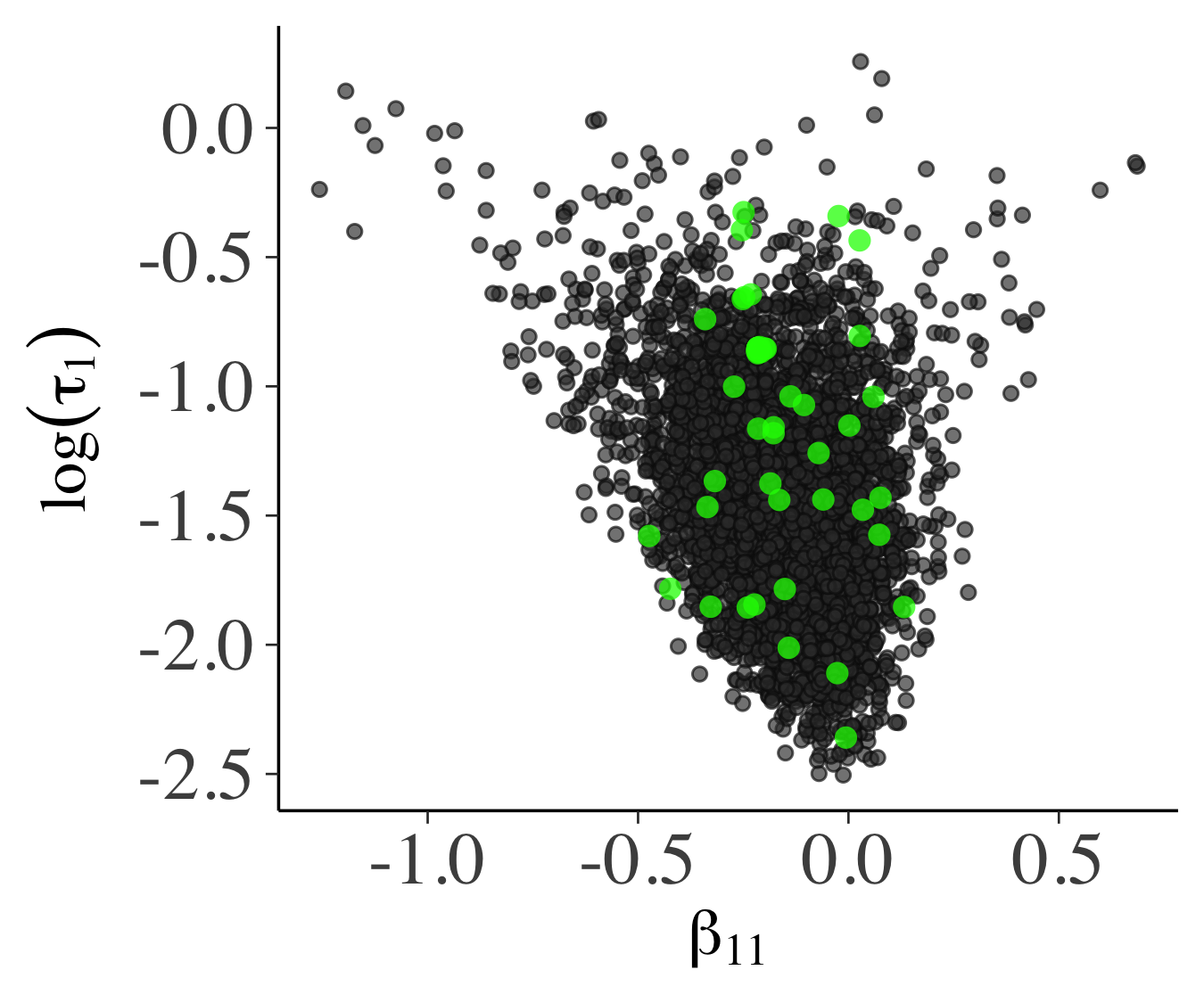}
\caption{For Model 3, a bivariate plot of the log standard deviation of the
cluster-level slopes ($y$-axis) against the slope for the first cluster
($x$-axis). The green dots indicate starting points of divergent transitions.
This plot can be made using {\tt mcmc\_scatter} in \bayesplot. }
\label{fig:mcmc_scatter_divs}
\end{subfigure}
~
\begin{subfigure}{0.48\textwidth}
\includegraphics[width=\textwidth]{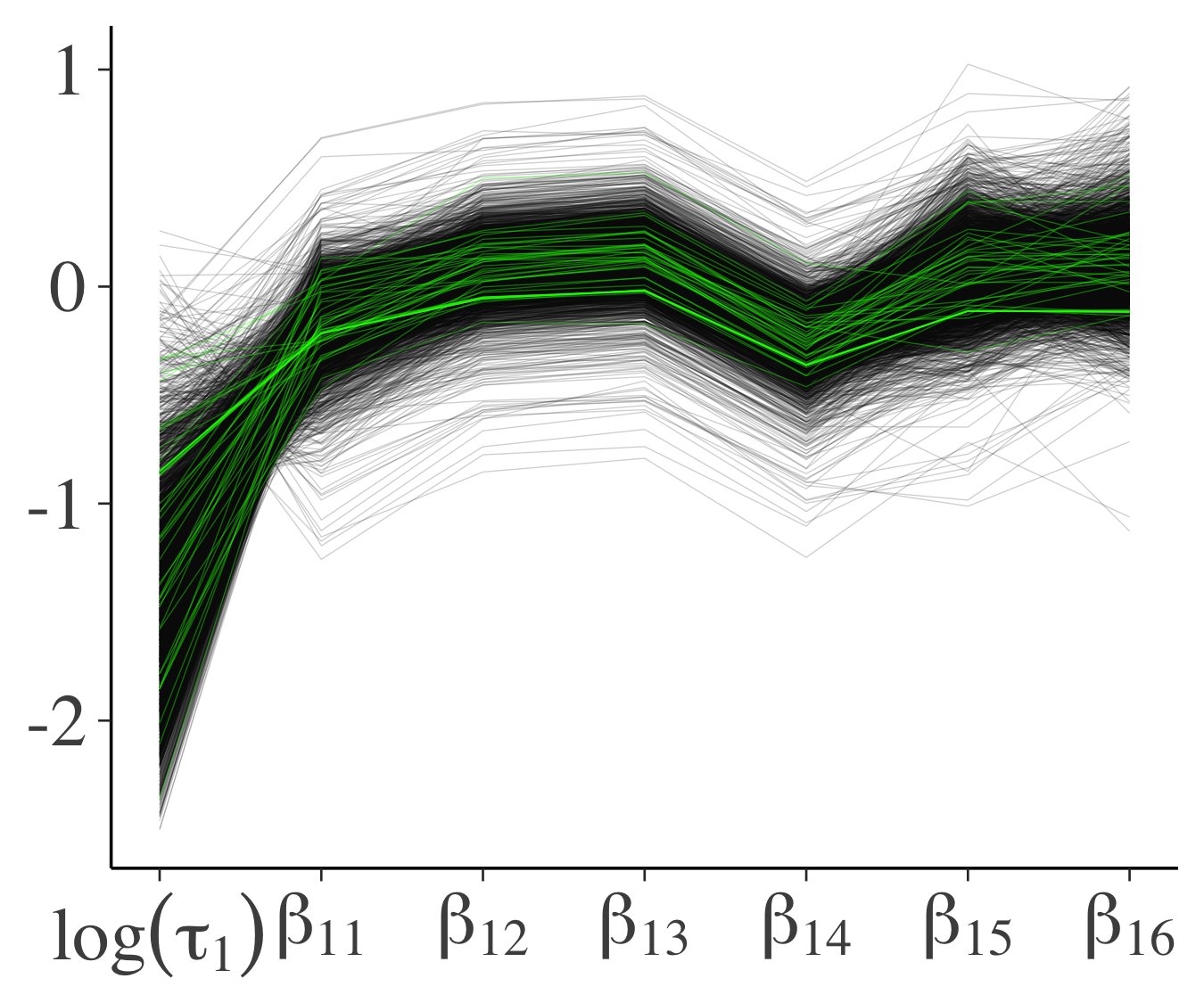}
\caption{For Model 3, a parallel coordinates plot showing the cluster-level
slope parameters and their log standard deviation $\log{\tau_1}$. The green
lines indicate starting points of divergent transitions. This plot can be made
using {\tt mcmc\_parcoord} in \bayesplot.}
\label{fig:mcmc_parcoord_divs}
\end{subfigure}

\caption{\it Diagnostic plots for Hamiltonian Monte Carlo.
Models were fit using the RStan interface to Stan 2.17 \citep{rstan}.}
\end{figure}

There are several plots that we have found useful for diagnosing troublesome
areas of the parameter space, in particular bivariate scatterplots that mark the
divergent transitions (Figure~\ref{fig:mcmc_scatter_divs}), and parallel
coordinate plots (Figure~\ref{fig:mcmc_parcoord_divs}). These visualizations are
sensitive enough to differentiate between models with a non-smooth typical set
and models where the heuristic has given a false positive. This makes them an
indispensable tool for understanding the behavior of a HMC when applied to a
particular target distribution.

If HMC were struggling to fit Model 3, the divergences would be clustered in the
parameter space. Examining the bivariate scatterplots
(Figure~\ref{fig:mcmc_scatter_divs}), there is no obvious pattern to the
divergences. Similarly, the parallel coordinate plot
(Figure~\ref{fig:mcmc_parcoord_divs}) does not show any particular structure.
This indicates that the divergences that are found are most likely false
positives.  For contrast, the supplementary material contains the same plots for
a model where HMC fails to compute a reliable answer. In this case, the
clustering of divergences is pronounced and the parallel coordinate plot clearly
indicates that all of the divergent trajectories have the same structure.

\section{How did we do? Posterior predictive checks are vital for model evaluation}
\label{sec:ppc}

\begin{figure}
\centering
\begin{subfigure}{0.31\textwidth}
\includegraphics[width=\textwidth]{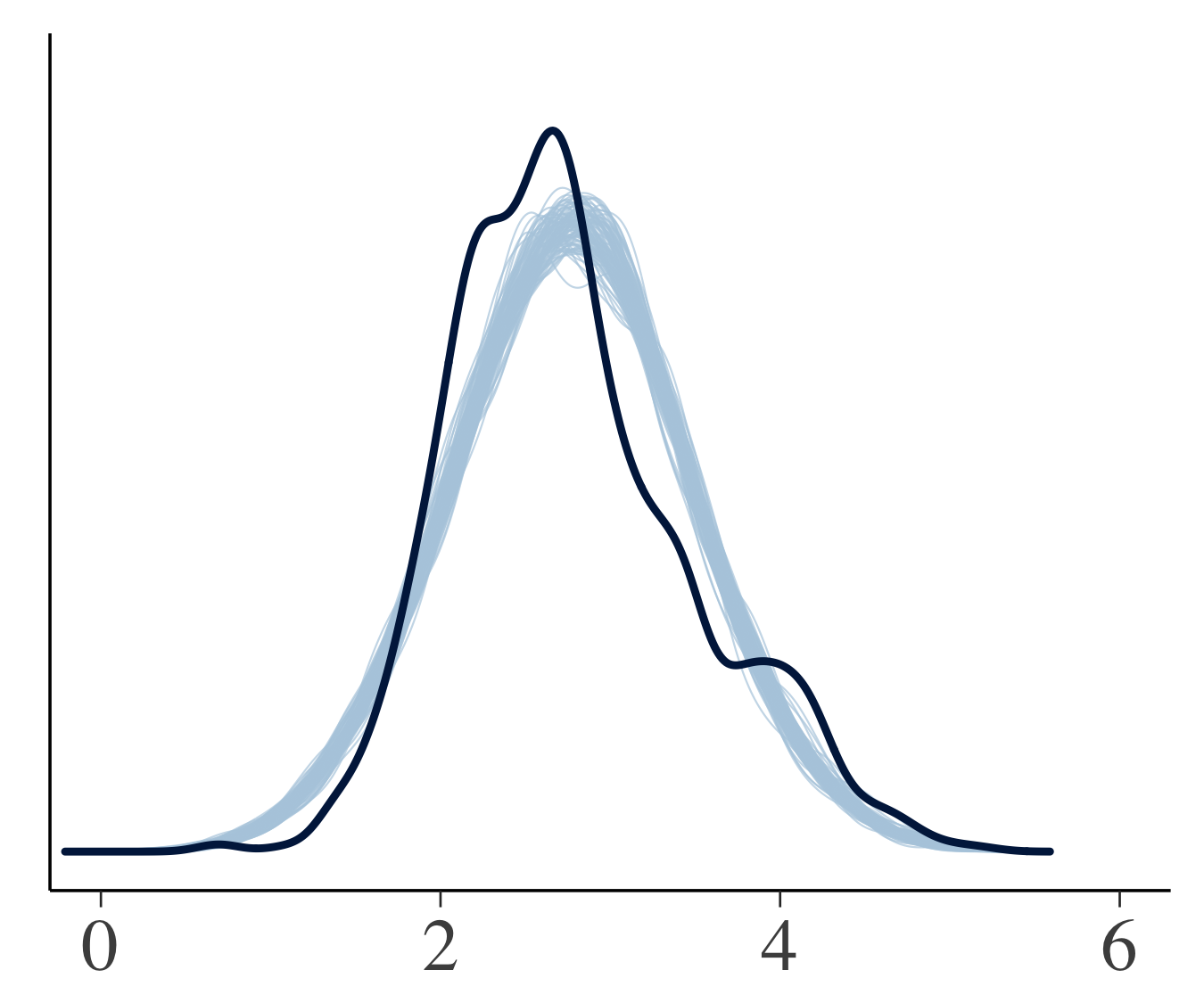}
\caption{Model 1}
\label{fig:ppc_dens0}
\end{subfigure}
~
\begin{subfigure}{0.31\textwidth}
\includegraphics[width=\textwidth]{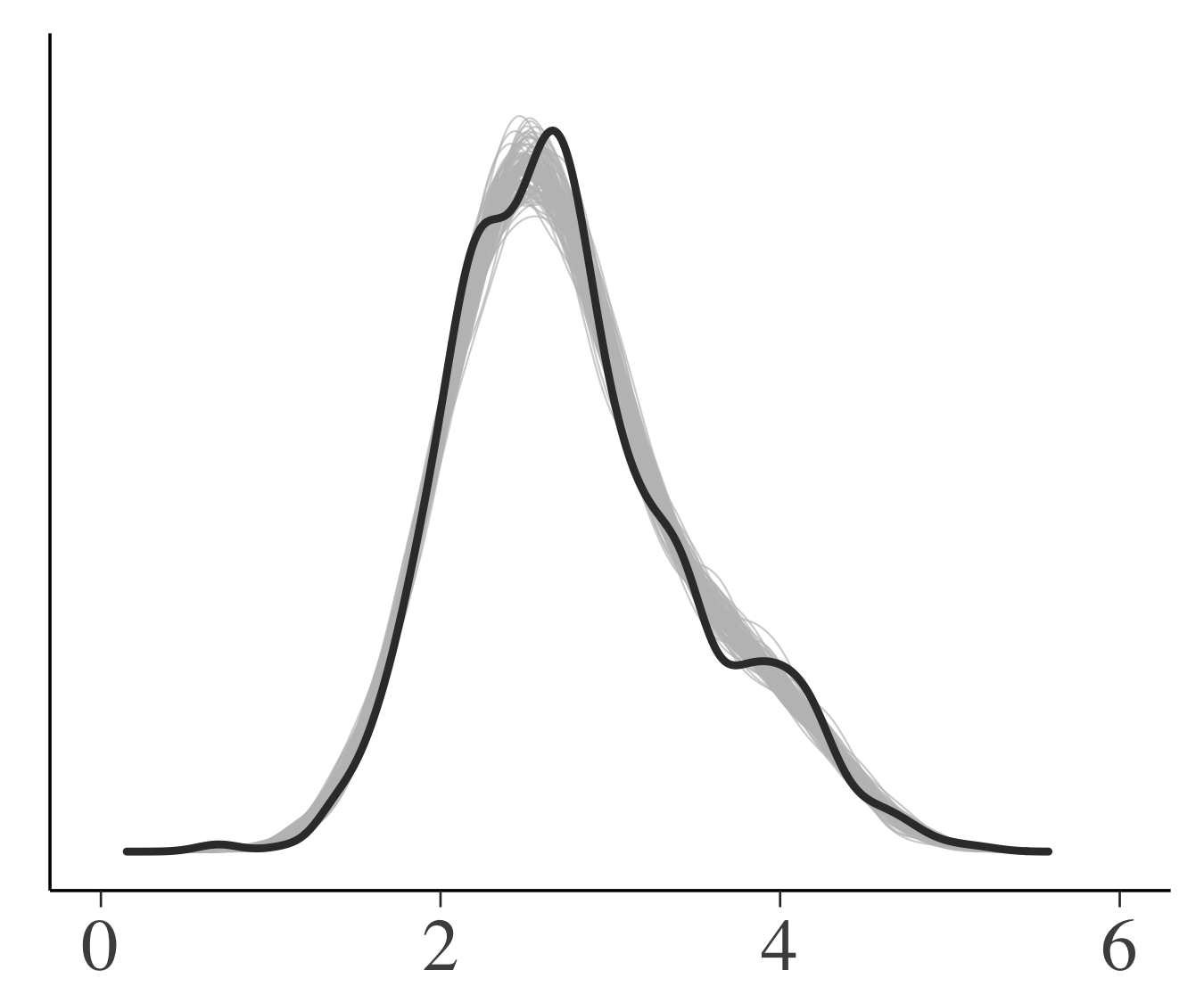}
\caption{Model 2}
\label{fig:ppc_dens2}
\end{subfigure}
~
\begin{subfigure}{0.31\textwidth}
\includegraphics[width=\textwidth]{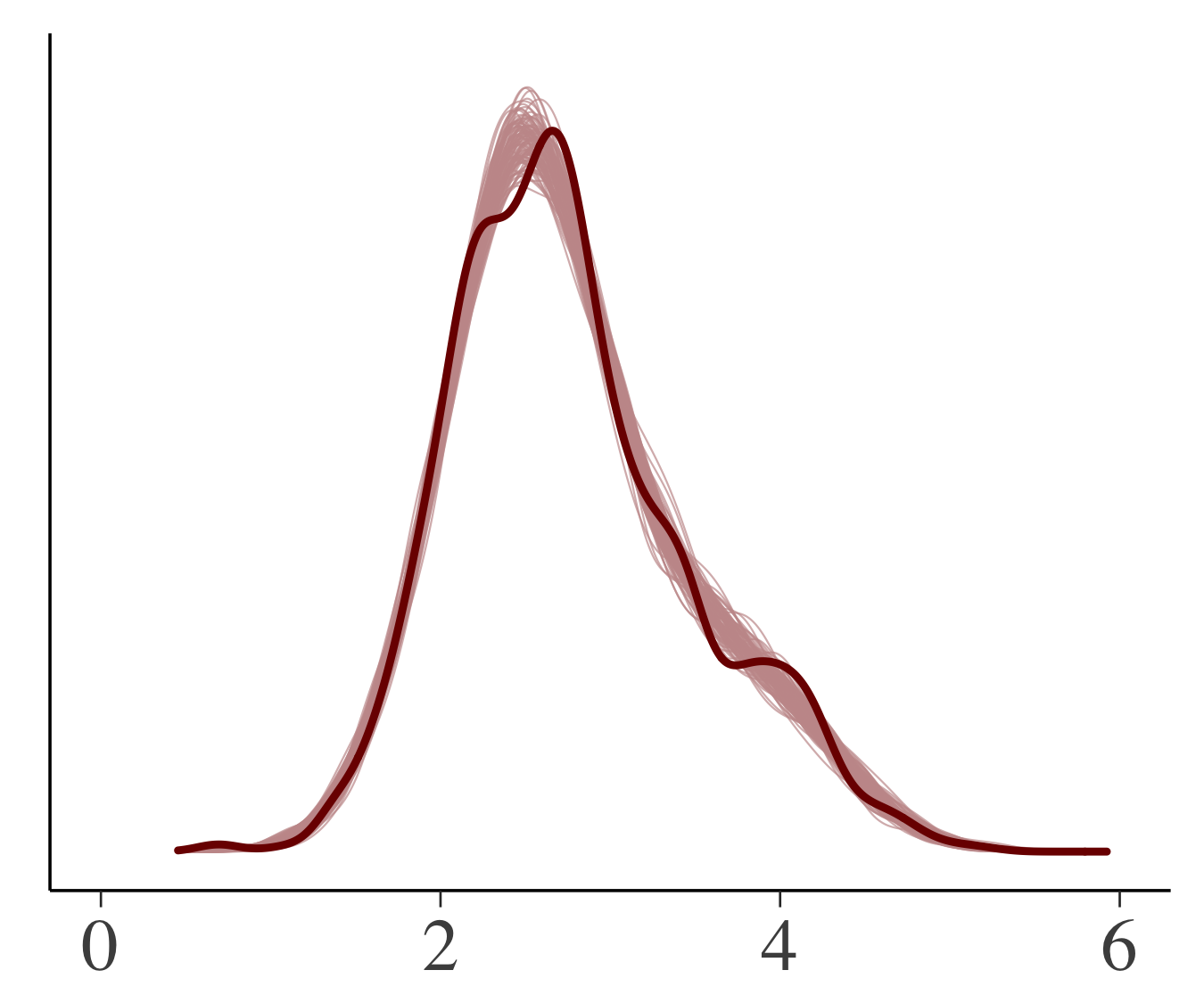}
\caption{Model 3}
\label{fig:ppc_dens3}
\end{subfigure}

\caption{\it Kernel density estimate of the observed dataset $y$ (dark curve),
with density estimates for 100 simulated datasets $y_{\rm rep}$ drawn
from the posterior predictive distribution (thin, lighter lines).
These plots can be produced using {\tt ppc\_dens\_overlay}
in the \bayesplot\ package.}
\label{fig:ppc_dens}
\end{figure}

The idea behind posterior predictive checking is simple: if a model is a good
fit  we should be able to use it to generate data that resemble the data we
observed. This is similar in spirit to the prior checks considered in Section
\ref{sec:prior_pred}, except now we have a data-informed data generating model.
This means we can be much more stringent in our comparisons. Ideally, we would
compare the model predictions to an independent test data set, but this is not
always feasible. However, we can still do some checking and predictive
performance assessments using the data we already have.

To generate the data used for posterior predictive checks (PPCs) we simulate
from the posterior predictive distribution $ p(\tilde{y} \mid y) = \int
p(\tilde{y} \mid \theta) p(\theta \mid y) \,d\theta, $ where $y$ is our current
data, $\tilde{y}$ is our new data to be predicted, and $\theta$ are our model
parameters. Posterior predictive checking is mostly qualitative. By looking at
some important features of the data and the replicated data, which were not
explicitly included in the model, we may find a need to extend or modify the
model.

For each of the three models, Figure~\ref{fig:ppc_dens} shows the distributions
of many replicated datasets drawn from the posterior predictive distribution
(thin light lines) compared to the empirical distribution of the observed
outcome (thick dark line). From these plots it is evident that the multilevel
models (2 and 3) are able to simulate new data that is more similar to the
observed $\log{({\rm PM}_{2.5})}$ values than the model without any hierarchical
structure (Model 1).

\begin{figure}
\centering
\begin{subfigure}{0.31\textwidth}
\includegraphics[width=\textwidth]{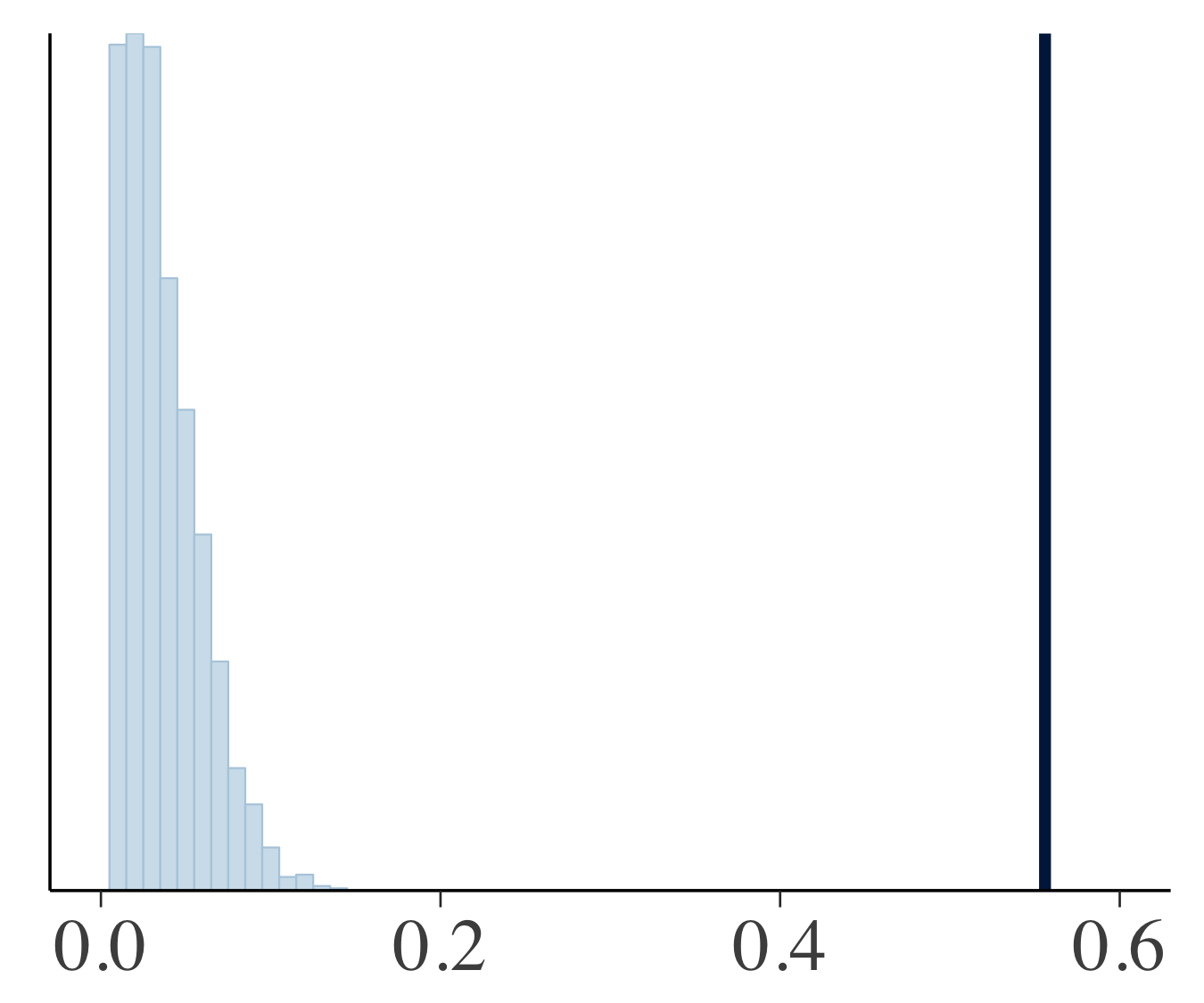}
\caption{Model 1}
\label{fig:ppc_skew0}
\end{subfigure}
~
\begin{subfigure}{0.31\textwidth}
\includegraphics[width=\textwidth]{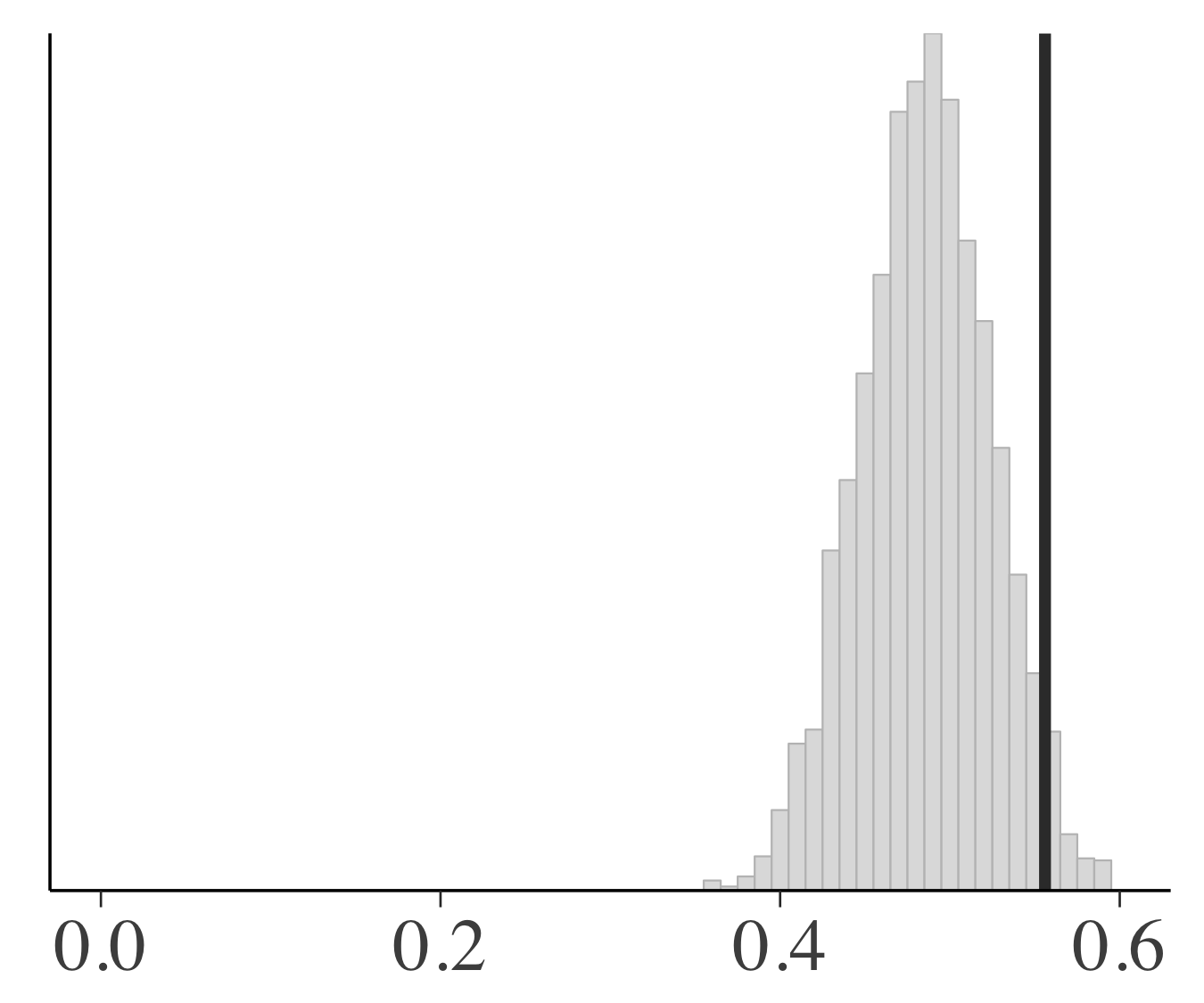}
\caption{Model 2}
\label{fig:ppc_skew2}
\end{subfigure}
~
\begin{subfigure}{0.31\textwidth}
\includegraphics[width=\textwidth]{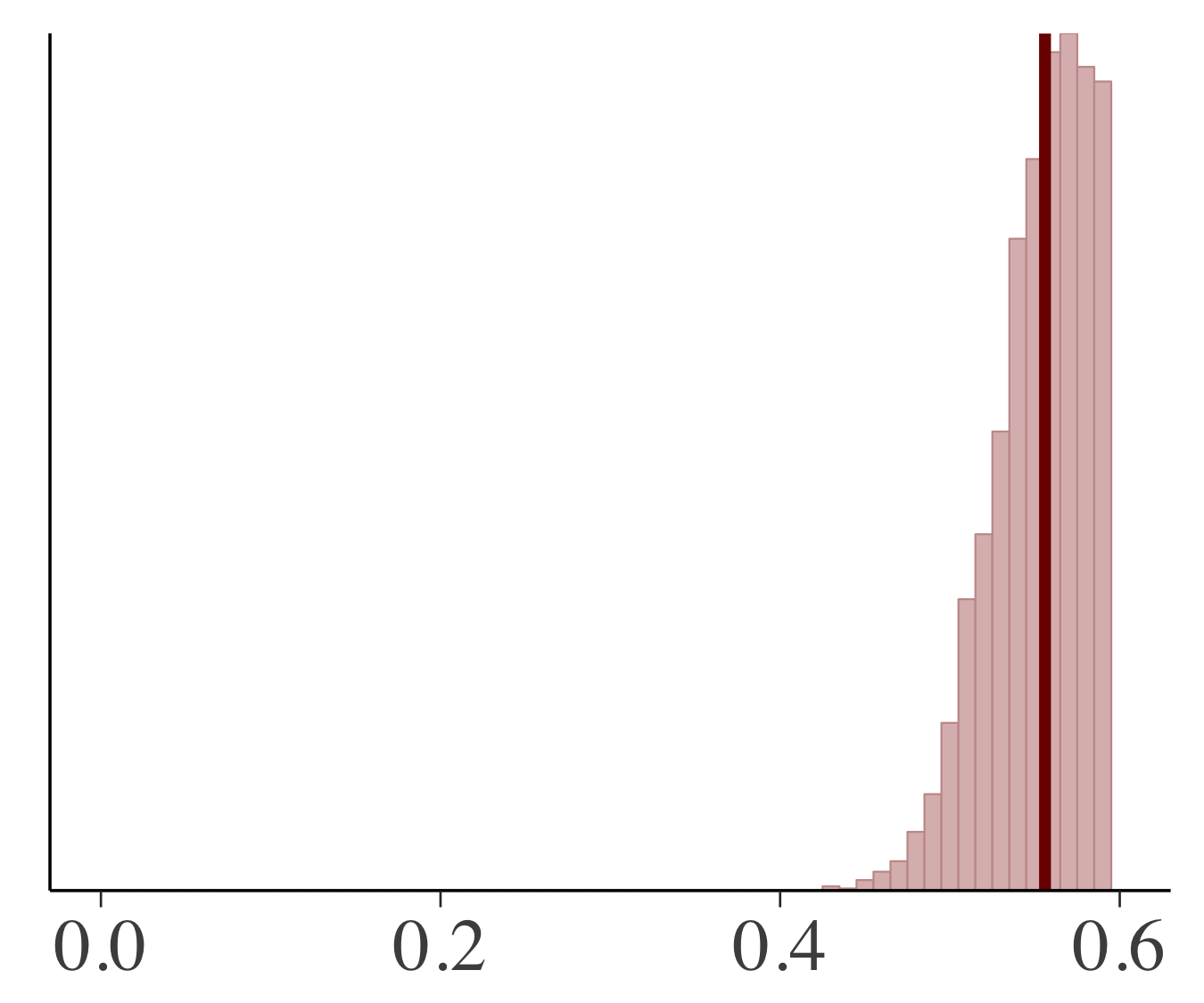}
\caption{Model 3}
\label{fig:ppc_skew3}
\end{subfigure}

\caption{\it Histograms of statistics ${\rm skew}{(y_{\rm rep})}$
computed from 4000 draws from the posterior predictive distribution. The dark
vertical line is computed from the observed data. These plots can be produced
using {\tt ppc\_stat} in the \bayesplot\ package.}
\label{fig:ppc_skew}
\end{figure}

Posterior predictive checking makes use of the data twice, once for the fitting
and once for the checking. Therefore it is a good idea to choose statistics that
are orthogonal to the model parameters. If the test statistic is related to one
of the model parameters, e.g., if the mean statistic is used for a Gaussian
model with a location parameter, the posterior predictive checks may be less
able to detect conflicts between the data and the model.
Our running example uses a Gaussian model so in Figure~\ref{fig:ppc_skew} we
investigate how well the posterior predictive distribution captures skewness.
Model 3, which used data-adapted regions, is best at capturing the observed
skewness, while Model 2 does an ok job and the linear regression (Model 1)
totally fails.

\begin{figure}
\centering
\begin{subfigure}{\textwidth}
\includegraphics[width=\textwidth]{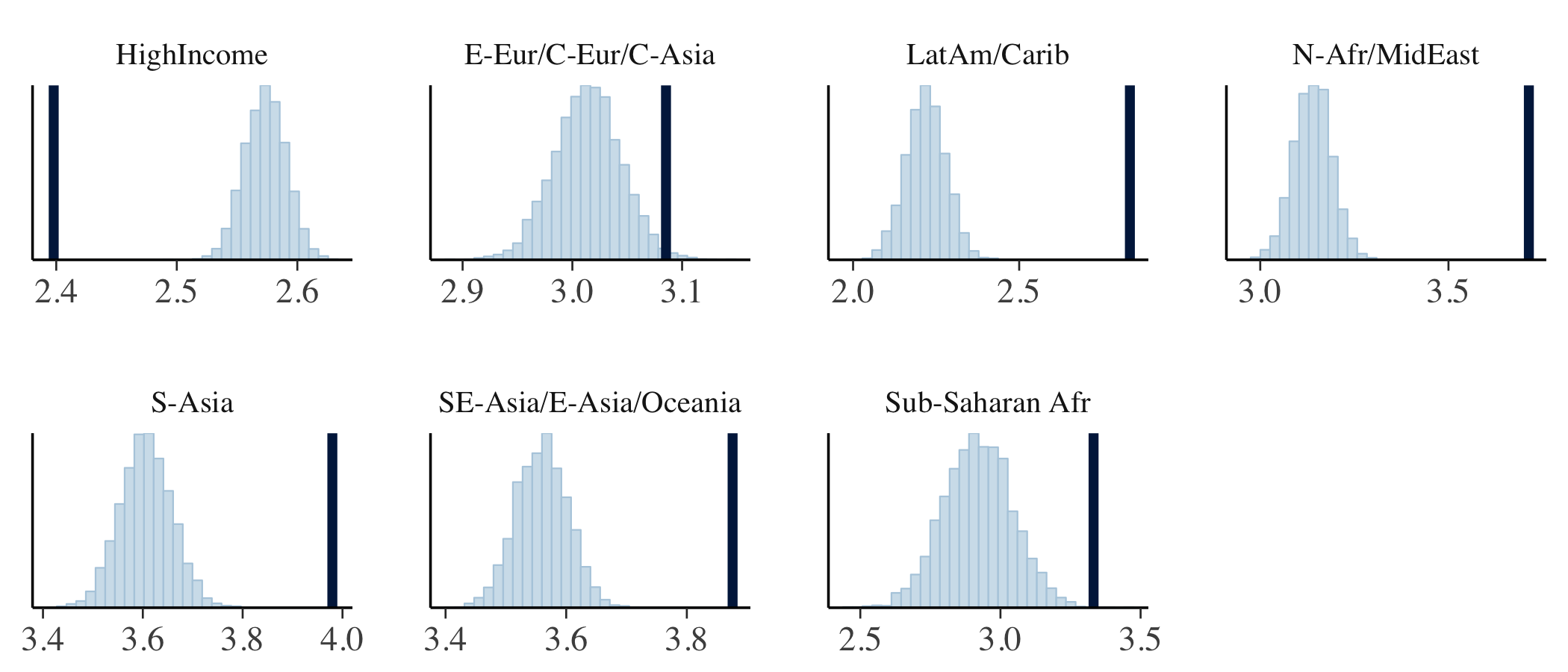}
\caption{Model 1}
\label{fig:ppc_med_grouped0}
\end{subfigure}
\begin{subfigure}{\textwidth}
\includegraphics[width=\textwidth]{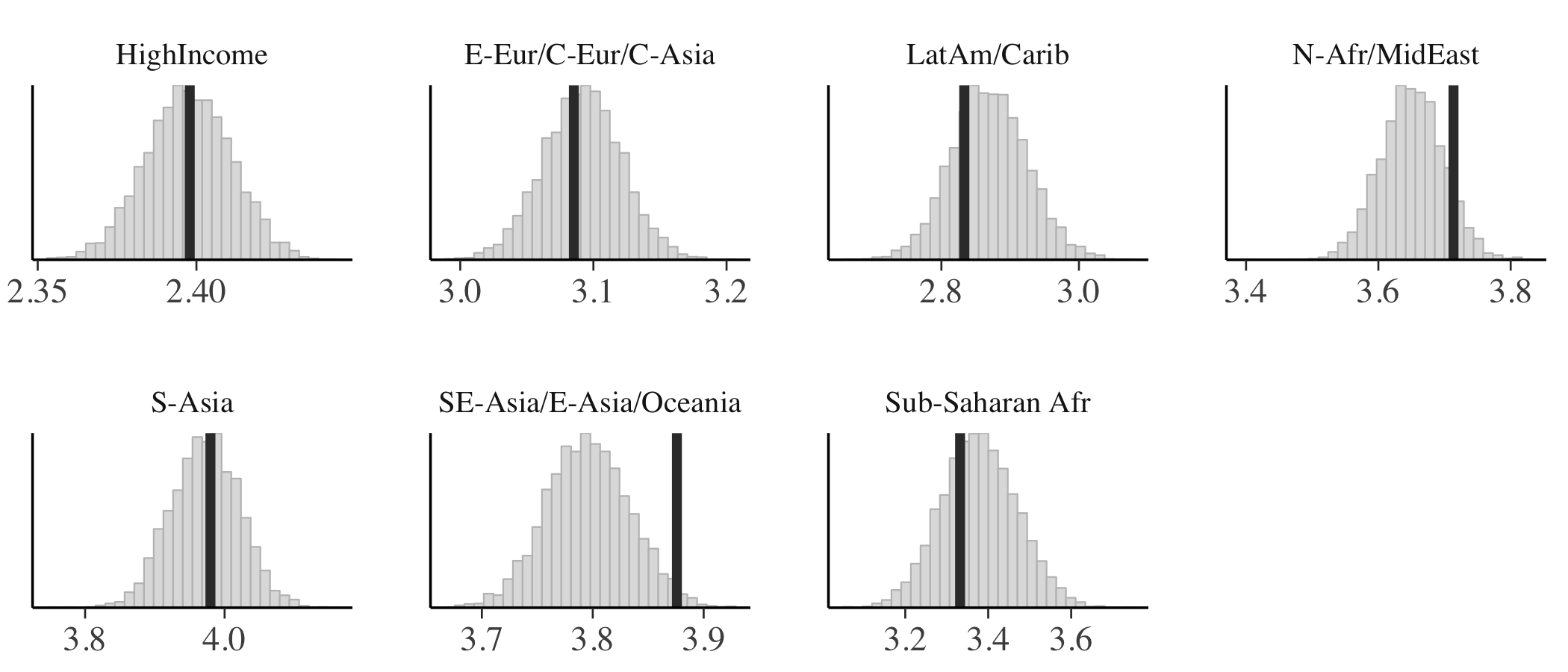}
\caption{Model 2}
\label{fig:ppc_med_grouped2}
\end{subfigure}
\begin{subfigure}{\textwidth}
\includegraphics[width=\textwidth]{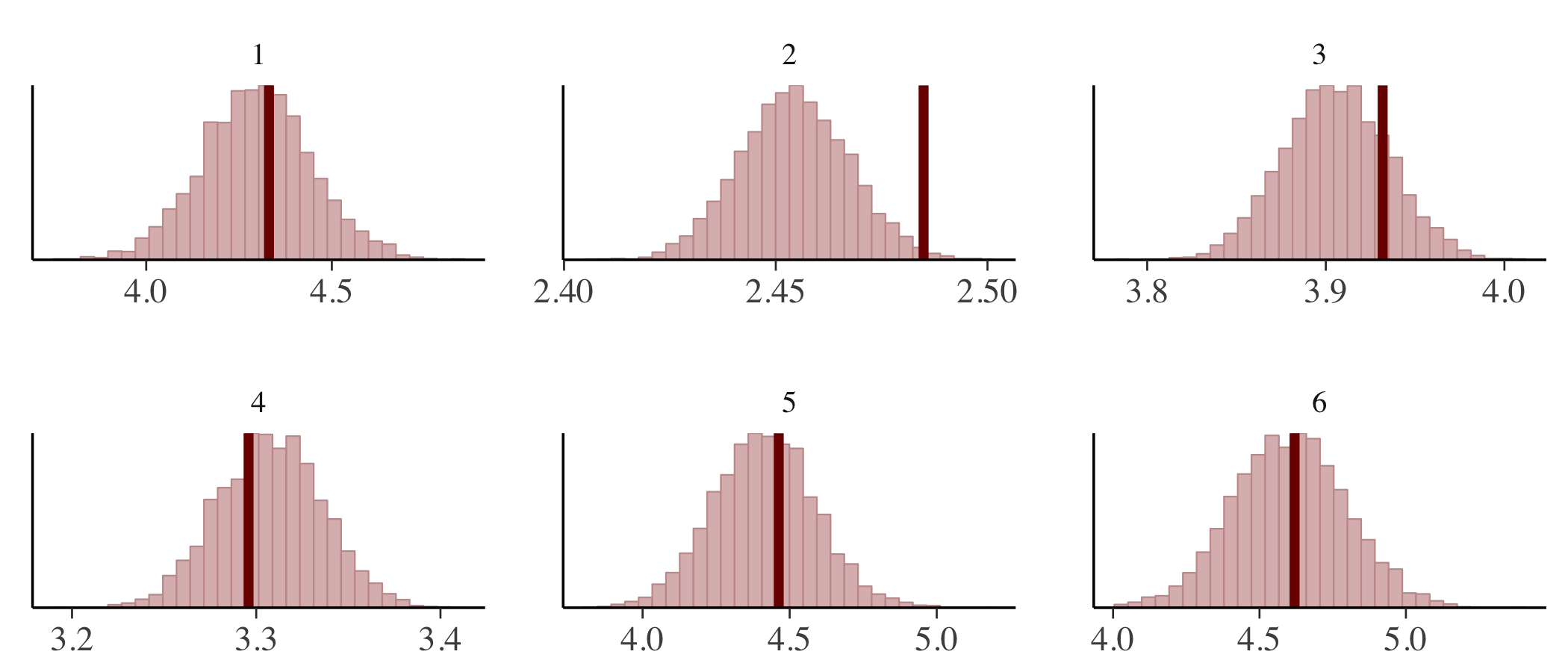}
\caption{Model 3}
\label{fig:ppc_med_grouped3}
\end{subfigure}

\caption{\it Checking posterior predictive test statistics, in this case the
medians, within region. The vertical lines are the observed medians. The facets
are labelled by number in panel (c) because they represent groups found by the
clustering algorithm rather than actual super-regions. These grouped plots can
be made using {\tt ppc\_stat\_grouped} in the \bayesplot\ package.}
\label{fig:ppc_med_grouped}
\end{figure}

We can also perform similar checks within levels of a grouping variable. For
example, in Figure~\ref{fig:ppc_med_grouped} we split both the outcome and
posterior predictive distribution according to region and check the median
values. The two hierarchical models give a better fit to the data at
the group level, which in this case is unsurprising.

In cross-validation, double use of data is partially avoided and test
statistics can be better calibrated.  When performing leave-one-out (LOO)
cross-validation we usually work with univariate posterior predictive
distributions, and thus we can't examine properties of the joint predictive
distribution. To specifically check that predictions are calibrated, the usual
test is to look at the leave-one-out cross-validation predictive cumulative
density function values, which are asymptotically uniform (for continuous data)
if the model is calibrated \citep{GelfandDeyChang1992, gelman2013bda}.

\begin{figure}
\centering
\begin{subfigure}{0.31\textwidth}
\includegraphics[width=\textwidth]{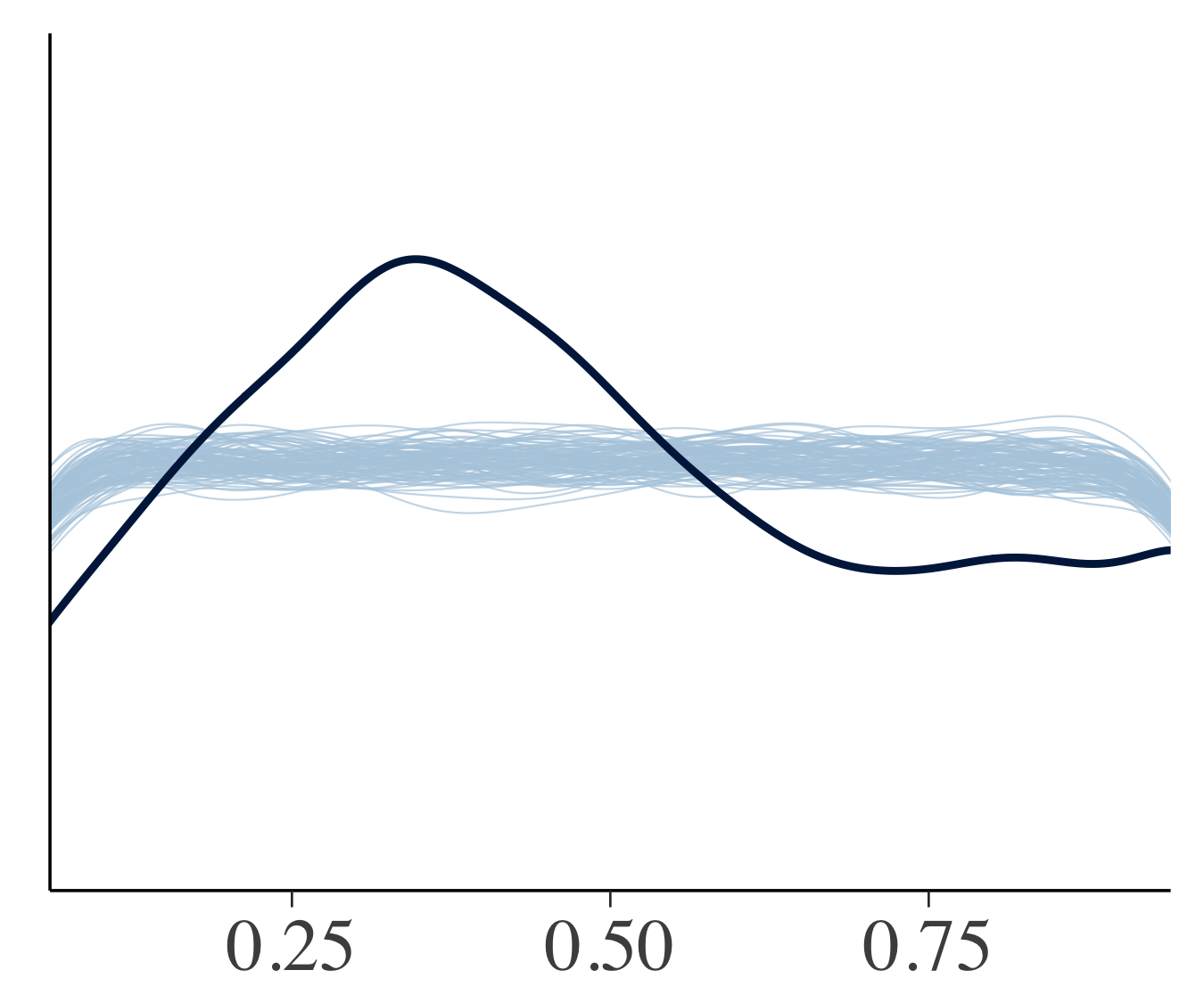}
\caption{Model 1}
\label{fig:ppc_loo_pit_unif0}
\end{subfigure}
~
\begin{subfigure}{0.31\textwidth}
\includegraphics[width=\textwidth]{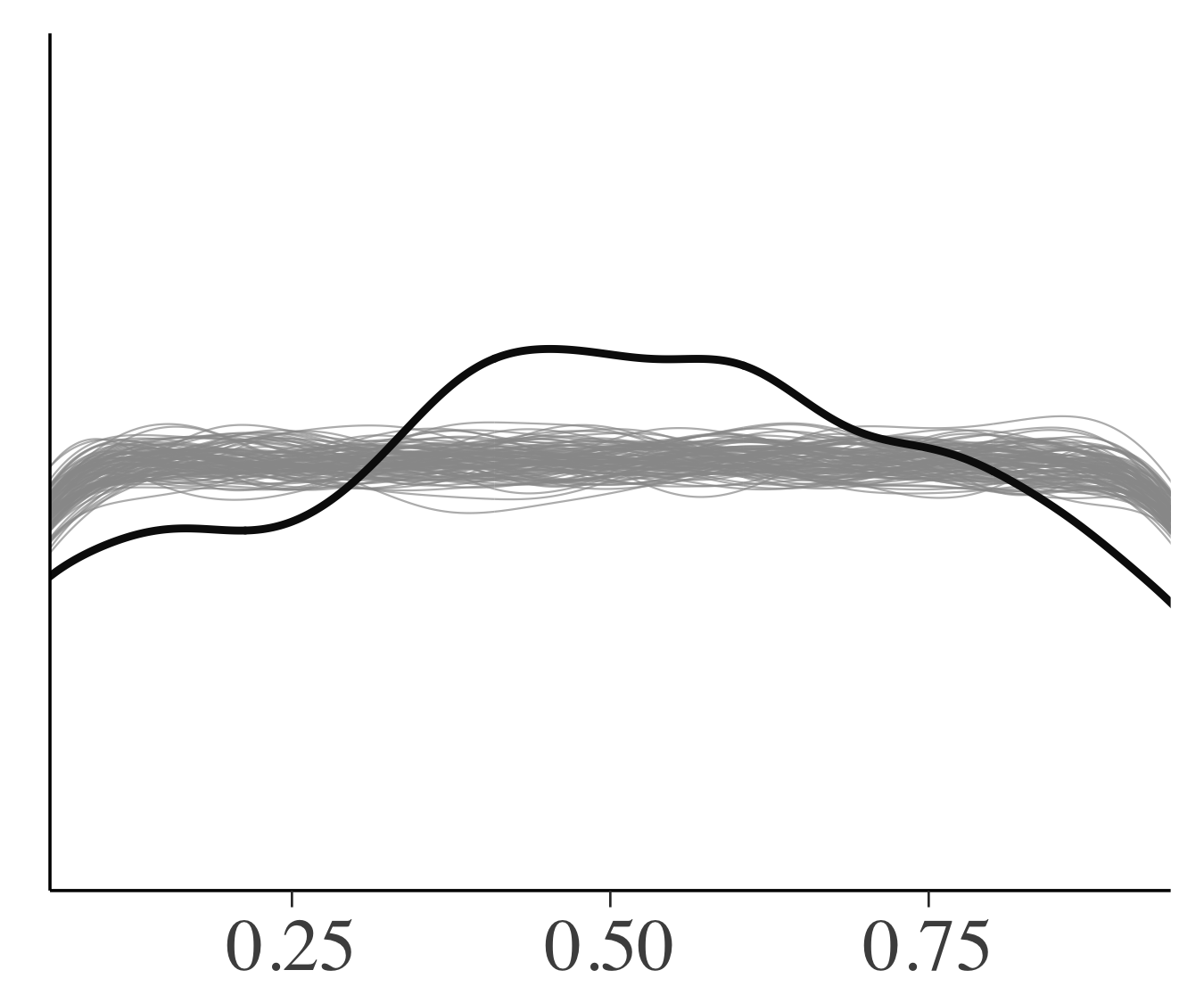}
\caption{Model 2}
\label{fig:ppc_loo_pit_unif2}
\end{subfigure}
~
\begin{subfigure}{0.31\textwidth}
\includegraphics[width=\textwidth]{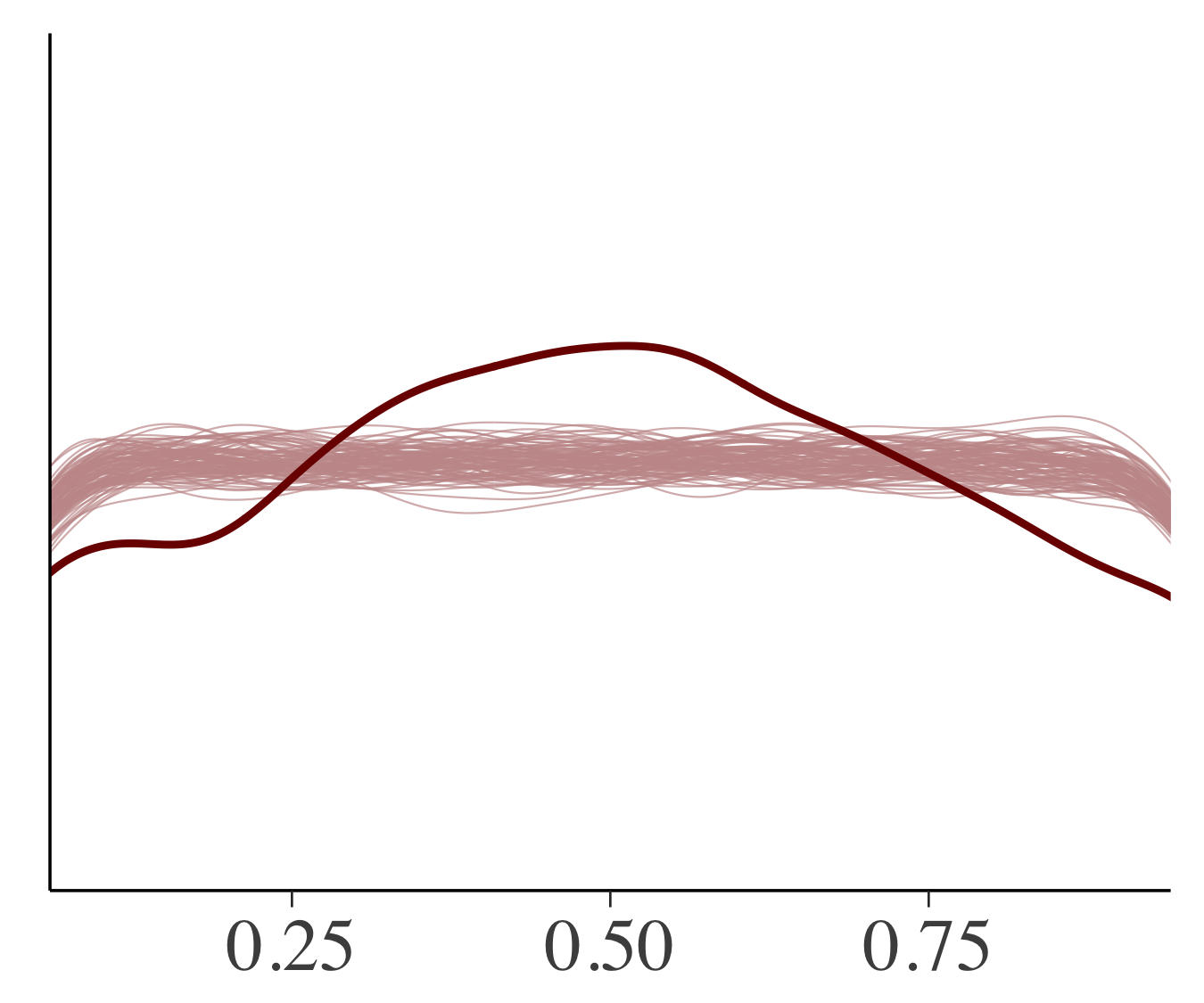}
\caption{Model 3}
\label{fig:ppc_loo_pit_unif3}
\end{subfigure}

\caption{\it Graphical check of leave-one-out cross-validated probability
integral transform (LOO-PIT). The thin lines represent simulations from the
standard uniform distribution and the thick dark line in each plot is the
density of the computed LOO-PITs. Similar plots can be made using
{\tt ppc\_dens\_overlay} and {\tt ppc\_loo\_pit}  in the \bayesplot\ package. 
The downwards slope near zero and one on the ``uniform" histograms is an 
edge effect due to the density estimator used and can be safely discounted.}
\label{fig:ppc_loo_pit_unif}
\end{figure}

The plots shown in Figure~\ref{fig:ppc_loo_pit_unif} compare the density of the
computed LOO probability integral transforms (thick dark line) versus 100 simulated 
datasets from a standard uniform distribution (thin light lines).  We can see that, 
although there is some clear miscalibration in all cases, the hierarchical models are 
an improvement over the single-level model.

The shape of the miscalibration in Figure~\ref{fig:ppc_loo_pit_unif} is also
meaningful. The frown shapes exhibited by Models 2 and 3 indicate that the
univariate predictive distributions are too broad compared to the data, which
suggests that further modeling will be necessary to accurately reflect the
uncertainty. One possibility would be to further sub-divide the super-regions to
better capture within-region variability \citep{shaddick2017data}.

\section{Pointwise plots for predictive model comparison}
Visual posterior predictive checks are also useful for identifying unusual
points in the data. Unusual data points come in two flavors: outliers and points
with high leverage.  In this section, we show that visualization can be useful
for identifying both types of data point. Examining these unusual observations
is a critical part of any statistical workflow, as these observations give hints
as to how the model may need to be modified. For example, they may indicate the
model should use non-linear instead of linear regression, or that the
observation error should be modeled with a heavier tailed distribution.

\begin{figure}
\centering
\begin{subfigure}{0.48\textwidth}
\includegraphics[width=\textwidth]{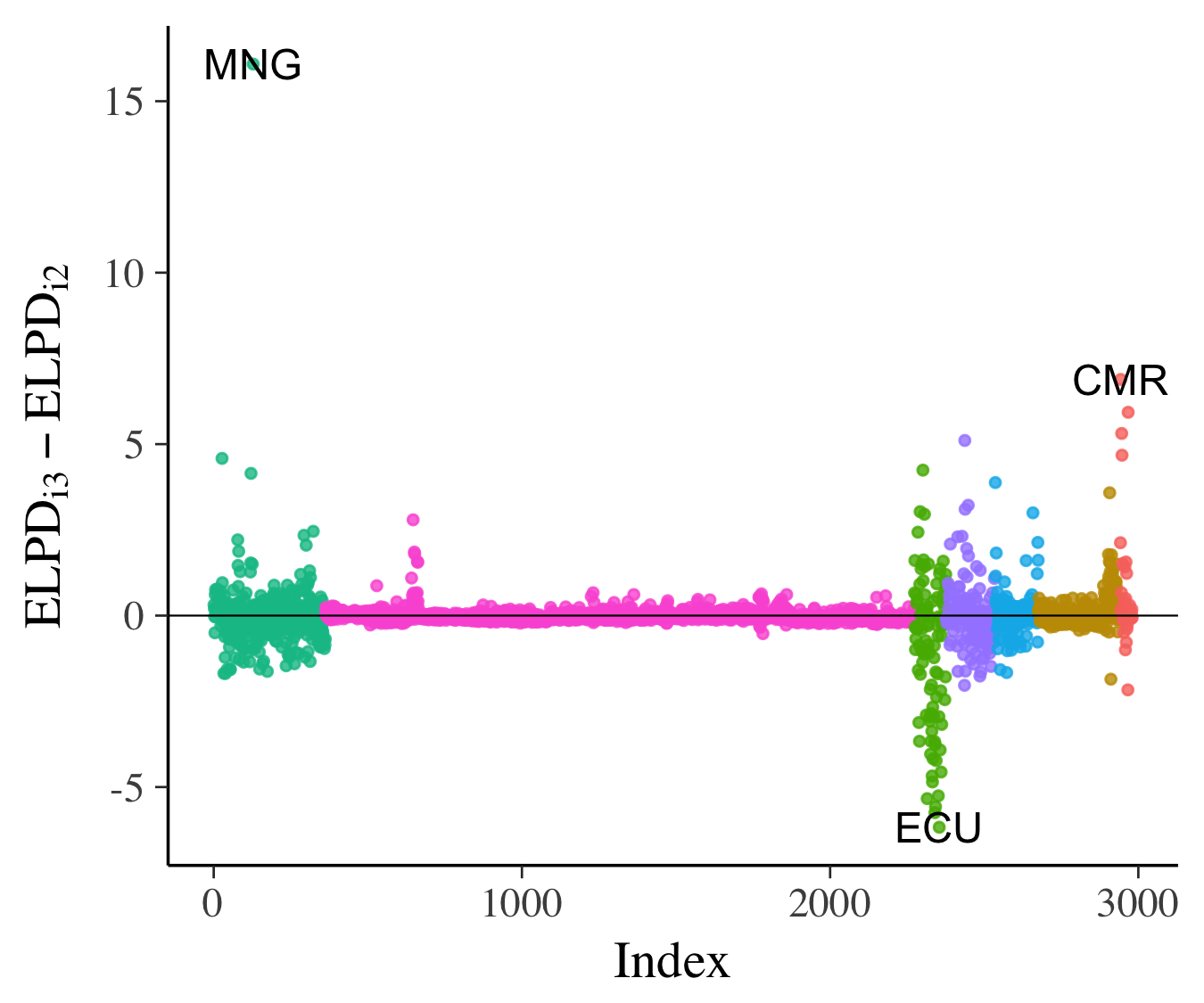}
\caption{The difference in pointwise ELPD values obtained from PSIS-LOO for
Model 3 compared to Model 2 colored by the WHO cluster (see
Figure~\ref{fig:plot1} for the key). Positive values indicate Model 3
outperformed Model 2.}
\label{fig:loo_elpd_diff}
\end{subfigure}
~
\begin{subfigure}{0.48\textwidth}
\includegraphics[width=\textwidth]{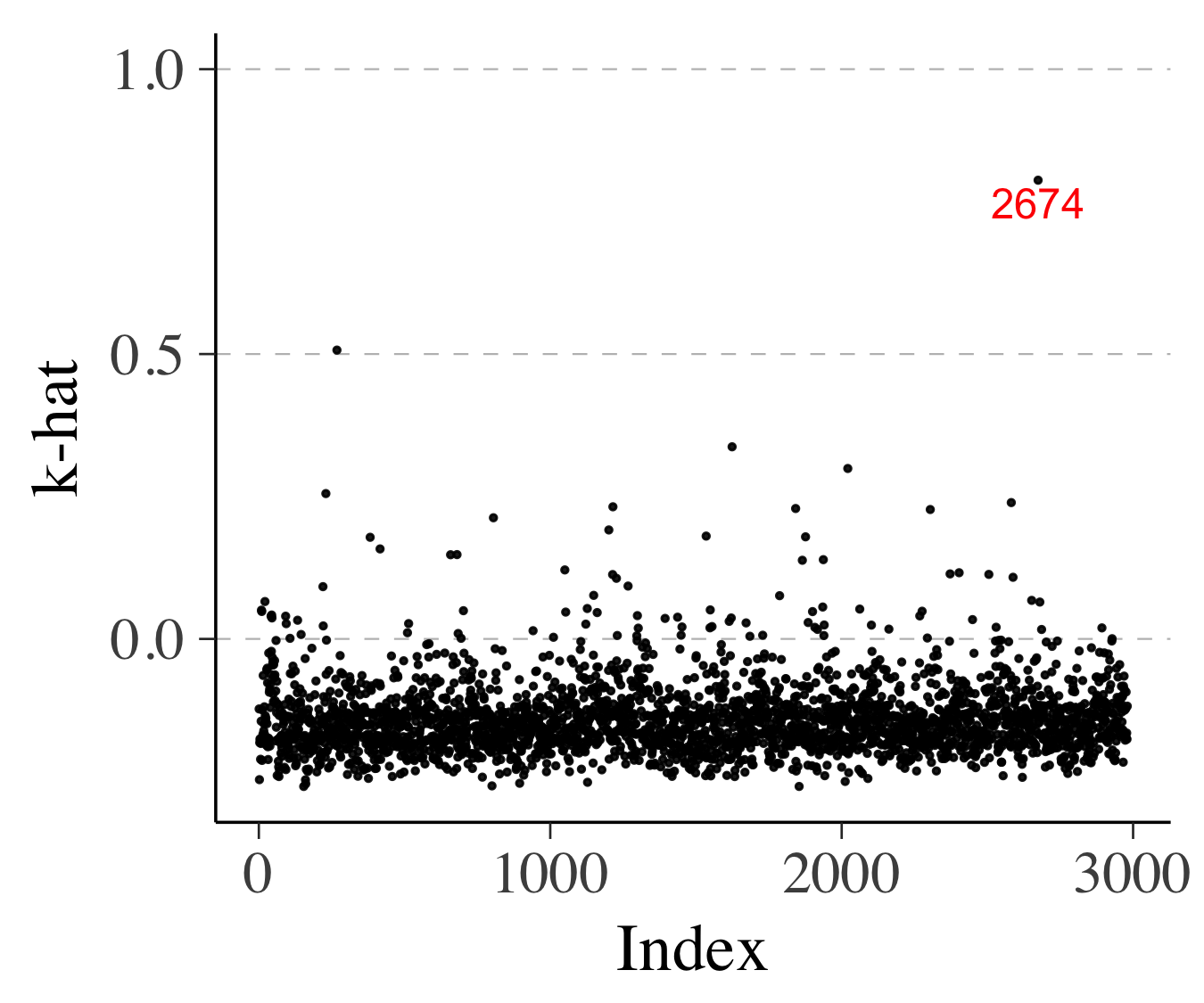}
\caption{The $\hat{k}$ diagnostics from PSIS-LOO for Model 2. The 2674th data
point (the only data point from Mongolia) is highlighted by the $\hat{k}$
diagnostic as being influential on the posterior.}
\label{fig:loo_khat}
\end{subfigure}
\caption{\it Model comparisons using leave-one-out (LOO) cross-validation.}
\end{figure}

The main tool in this section is the one-dimensional cross-validated
leave-one-out (LOO) predictive distribution $p(y_i \mid y_{-i})$.
\citet{GelfandDeyChang1992} suggested examining the LOO log-predictive density
values (they called them conditional predictive ordinates) to find observations
that are difficult to predict. This idea can be extended to model comparison by
looking at which model best captures each left out data point.
Figure~\ref{fig:loo_elpd_diff} shows the difference between the expected log
predictive densities (ELPD) for the individual data points estimated using
Pareto-smoothed importance sampling (PSIS-LOO, \citet{vehtari2016psis,
vehtari2017practical}). Model 3 appears to be slightly better than Model 2,
especially for difficult observations like the station in Mongolia.

In addition to looking at the individual LOO log-predictive densities, it is
useful to look at how influential each observation is. Some of the data points
may be difficult to predict but not necessarily influential, that is, the
predictive distribution does not change much when they are left out. 
One way to look at the influence is to look at the difference between the
full data log-posterior predictive density and the LOO log-predictive density.

We recommend computing the LOO log-predictive densities using PSIS-LOO as
implemented in the {\tt loo} package \citep{looRpackage}. A key advantage of
using PSIS-LOO to compute the LOO densities is that it automatically computes an
empirical estimate of how similar the full-data predictive distribution is to
the LOO predictive distribution for each left out point. Specifically, it
computes an empirical estimate $\hat{k}$ of
$k=\inf\left\{ k'>0 : D_{\frac{1}{k'}} \left(  p || q  \right)<\infty \right\},$
where
$D_{\alpha} \left(  p || q  \right) =
\frac{1}{\alpha-1} \log \int_{\Theta} p(\theta)^{\alpha} q(\theta)^{1-\alpha}  d \theta$
is the $\alpha$-R\'{e}nyi divergence \citep{yao2018yes}.
If the $j$th LOO predictive distribution has a large $\hat{k}$ value when used as
a proposal distribution for the full-data predictive distribution, it suggests
that $y_j$ is a highly influential observation.

Figure~\ref{fig:loo_khat} shows the $\hat{k}$ diagnostics from PSIS-LOO for our
Model 2. The 2674th data point is highlighted by the $\hat{k}$ diagnostic as
being influential on the posterior. If we examine the data we find that this
point is the only observation from Mongolia and corresponds to a measurement
$(x,y) = (\log{(\rm satellite)}, \log{({\rm PM}_{2.5})}) = (1.95, 4.32)$,
which would look extreme if highlighted in the scatterplot in Figure~\ref{fig:plot1}. 
By contrast, under Model 3 the $\hat{k}$ value for the Mongolian observation 
is significantly lower ($\hat{k} \approx 0.5$) indicating that the data point is better 
resolved in Model 3.

\section{Discussion}

Visualization is probably the most important tool in an applied statistician's
toolbox and is an important complement to quantitative statistical procedures \citep{buja2009statistical}.
In this paper, we've demonstrated that it can be used as part of a strategy to
compare models, to identify ways in which a model fails to fit, to check how
well our computational methods have resolved the model, to understand the model
well enough to be able to set priors, and to iteratively improve the model.

The last of these tasks is a little bit controversial as using the measured data
to guide model building raises the concern that the resulting model will
generalize poorly to new datasets. A different objection to using the data twice
(or even more) comes from ideas around hypothesis testing and unbiased
estimation, but we are of the opinion that the danger of overfitting the data is
much more concerning \citep{gelman2014statistical}.

In the visual workflow we've outlined in this paper, we have used the data to
improve the model in two places. In Section~\ref{sec:prior_pred} we proposed
prior predictive checks with the recommendation that the data generating
mechanism should be broader than the distribution of the observed data in line
with the principle of weakly informative priors. In Section~\ref{sec:ppc} we
recommended undertaking careful calibration checks as well as checks based on
summary statistics, and then updating the model accordingly to cover the
deficiencies exposed by this procedure. In both of these cases, we have made
recommendations that aim to reduce the danger. For the prior predictive checks,
we recommend not cleaving too closely to the observed data and instead aiming
for a prior data generating process that can produce plausible data sets, not
necessarily ones that are indistinguishable from observed data. For the
posterior predictive checks, we ameliorate the concerns by checking carefully
for influential measurements and proposing that model extensions be weakly
informative extensions that are still centered on the previous model
\citep{simpson2017penalising}.

Regardless of concerns we have about using the data twice, the workflow that we
have described in this paper (perhaps without the stringent prior and posterior
predictive checks) is common in applied statistics. As academic statisticians,
we have a duty to understand the consequences of this workflow and offer
concrete suggestions to make the practice of applied statistics more robust.

\subsection*{Acknowledgements}

The authors thank Gavin Shaddick and Matthew Thomas for their help with the \PM\
example, Ari Hartikainen for suggesting the parallel coordinates plot,
Ghazal Fazelnia for finding an error in our map of ground monitor locations,
Eren Metin El\c{c}i for alerting us to a discrepancy between our text and code,
and the Sloan Foundation, Columbia University, U.S. National Science Foundation,
Institute for Education Sciences, Office of Naval Research, and 
Defense Advanced Research Projects Agency for financial support.

\bibliographystyle{chicago}
\bibliography{bayes-vis}

\begin{thebibliography}{}

\bibitem[\protect\citeauthoryear{Betancourt}{Betancourt}{2017}]{betancourt2017conceptual}
Betancourt, M. (2017).
\newblock A conceptual introduction to {Hamiltonian Monte Carlo}.
\newblock arXiv preprint arxiv:1701.02434.

\bibitem[\protect\citeauthoryear{Betancourt and Girolami}{Betancourt and
  Girolami}{2015}]{betancourt2015}
Betancourt, M. and M.~Girolami (2015).
\newblock {Hamiltonian Monte Carlo} for hierarchical models.
\newblock In S.~K. Upadhyay, U.~Singh, D.~K. Dey, and A.~Loganathan (Eds.),
  {\em Current Trends in {Bayesian} Methodology with Applications}, pp.\
  79--101. Chapman \& Hall.
\newblock arXiv:1312.0906.

\bibitem[\protect\citeauthoryear{Buja, Cook, Hofmann, Lawrence, Lee, Swayne,
  and Wickham}{Buja et~al.}{2009}]{buja2009statistical}
Buja, A., D.~Cook, H.~Hofmann, M.~Lawrence, E.-K. Lee, D.~F. Swayne, and
  H.~Wickham (2009).
\newblock Statistical inference for exploratory data analysis and model
  diagnostics.
\newblock {\em Philosophical Transactions of the Royal Society of London A:
  Mathematical, Physical and Engineering Sciences\/}~{\em 367\/}(1906),
  4361--4383.

\bibitem[\protect\citeauthoryear{Forouzanfar, Alexander, Anderson, Bachman,
  Biryukov, Brauer, Burnett, Casey, Coates, Cohen, et~al.}{Forouzanfar
  et~al.}{2015}]{forouzanfar2015global}
Forouzanfar, M.~H., L.~Alexander, H.~R. Anderson, V.~F. Bachman, S.~Biryukov,
  M.~Brauer, R.~Burnett, D.~Casey, M.~M. Coates, A.~Cohen, et~al. (2015).
\newblock Global, regional, and national comparative risk assessment of 79
  behavioural, environmental and occupational, and metabolic risks or clusters
  of risks in 188 countries, 1990--2013: a systematic analysis for the {Global
  Burden of Disease Study 2013}.
\newblock {\em The Lancet\/}~{\em 386\/}(10010), 2287--2323.

\bibitem[\protect\citeauthoryear{Gabry}{Gabry}{2017}]{bayesplotRpackage}
Gabry, J. (2017).
\newblock bayesplot: Plotting for {Bayesian} models.
\newblock {R} package version 1.3.0, \url{http://mc-stan.org/bayesplot}.

\bibitem[\protect\citeauthoryear{Gelfand, Dey, and Chang}{Gelfand
  et~al.}{1992}]{GelfandDeyChang1992}
Gelfand, A.~E., D.~K. Dey, and H.~Chang (1992).
\newblock Model determination using predictive distributions with
  implementation via sampling-based methods (with discussion).
\newblock In J.~M. Bernardo, J.~O. Berger, A.~P. Dawid, and A.~F.~M. Smith
  (Eds.), {\em Bayesian Statistics 4}, pp.\  147--167. Oxford University Press.

\bibitem[\protect\citeauthoryear{Gelman}{Gelman}{2004}]{gelman2004exploratory}
Gelman, A. (2004).
\newblock Exploratory data analysis for complex models.
\newblock {\em Journal of Computational and Graphical Statistics\/}~{\em
  13\/}(4), 755--779.

\bibitem[\protect\citeauthoryear{Gelman, Carlin, Stern, Dunson, Vehtari, and
  Rubin}{Gelman et~al.}{2013}]{gelman2013bda}
Gelman, A., J.~B. Carlin, H.~S. Stern, D.~B. Dunson, A.~Vehtari, and D.~B.
  Rubin (2013).
\newblock {\em Bayesian Data Analysis\/} (Third ed.).
\newblock Chapman \& Hall/CRC.
\newblock Chapter 6, Section ``Marginal predictive checks".

\bibitem[\protect\citeauthoryear{Gelman, Jakulin, Pittau, Su, et~al.}{Gelman
  et~al.}{2008}]{gelman2008weakly}
Gelman, A., A.~Jakulin, M.~G. Pittau, Y.-S. Su, et~al. (2008).
\newblock A weakly informative default prior distribution for logistic and
  other regression models.
\newblock {\em The Annals of Applied Statistics\/}~{\em 2\/}(4), 1360--1383.

\bibitem[\protect\citeauthoryear{Gelman and Loken}{Gelman and
  Loken}{2014}]{gelman2014statistical}
Gelman, A. and E.~Loken (2014).
\newblock The statistical crisis in science: Data-dependent analysis-a ``garden
  of forking paths"-explains why many statistically significant comparisons
  don't hold up.
\newblock {\em American Scientist\/}~{\em 102\/}(6), 460.

\bibitem[\protect\citeauthoryear{Gelman, Simpson, and Betancourt}{Gelman
  et~al.}{2017}]{gelman2017priors}
Gelman, A., D.~Simpson, and M.~Betancourt (2017).
\newblock The prior can generally only be understood in the context of the
  likelihood.
\newblock {\em arXiv preprint arXiv:1708.07487\/}.

\bibitem[\protect\citeauthoryear{{R Core Team}}{{R Core Team}}{2017}]{rcore}
{R Core Team} (2017).
\newblock {\em {R}: {A} Language and Environment for Statistical Computing}.
\newblock Vienna, Austria: R Foundation for Statistical Computing.

\bibitem[\protect\citeauthoryear{Rubin}{Rubin}{1981}]{rubin1981}
Rubin, D.~B. (1981).
\newblock Estimation in parallel randomized experiments.
\newblock {\em Journal of Educational Statistics\/}~{\em 6}, 377--401.

\bibitem[\protect\citeauthoryear{Shaddick, Thomas, Green, Brauer, Donkelaar,
  Burnett, Chang, Cohen, Dingenen, Dora, et~al.}{Shaddick
  et~al.}{2017}]{shaddick2017data}
Shaddick, G., M.~L. Thomas, A.~Green, M.~Brauer, A.~Donkelaar, R.~Burnett,
  H.~H. Chang, A.~Cohen, R.~V. Dingenen, C.~Dora, et~al. (2017).
\newblock Data integration model for air quality: a hierarchical approach to
  the global estimation of exposures to ambient air pollution.
\newblock {\em Journal of the Royal Statistical Society: Series C (Applied
  Statistics)\/}~{\em Available online 13 June 2017}.
\newblock arXiv:1609.00141.

\bibitem[\protect\citeauthoryear{Simpson, Rue, Riebler, Martins, S{\o}rbye,
  et~al.}{Simpson et~al.}{2017}]{simpson2017penalising}
Simpson, D., H.~Rue, A.~Riebler, T.~G. Martins, S.~H. S{\o}rbye, et~al. (2017).
\newblock Penalising model component complexity: A principled, practical
  approach to constructing priors.
\newblock {\em Statistical science\/}~{\em 32\/}(1), 1--28.

\bibitem[\protect\citeauthoryear{{Stan Development Team}}{{Stan Development
  Team}}{2017a}]{rstan}
{Stan Development Team} (2017a).
\newblock {RStan}: the {R} interface to {S}tan, version 2.16.1.
\newblock \url{http://mc-stan.org}.

\bibitem[\protect\citeauthoryear{{Stan Development Team}}{{Stan Development
  Team}}{2017b}]{stanmanual}
{Stan Development Team} (2017b).
\newblock {\em Stan Modeling Language User's Guide and Reference Manual,
  Version 2.16.0}.
\newblock \url{http://mc-stan.org}.

\bibitem[\protect\citeauthoryear{Vehtari, Gelman, and Gabry}{Vehtari
  et~al.}{2017a}]{looRpackage}
Vehtari, A., A.~Gelman, and J.~Gabry (2017a).
\newblock loo: {E}fficient leave-one-out cross-validation and {WAIC} for
  {B}ayesian models.
\newblock {R} package version 1.0.0, \url{http://mc-stan.org/loo}.

\bibitem[\protect\citeauthoryear{Vehtari, Gelman, and Gabry}{Vehtari
  et~al.}{2017b}]{vehtari2016psis}
Vehtari, A., A.~Gelman, and J.~Gabry (2017b).
\newblock Pareto smoothed importance sampling.
\newblock arXiv preprint arXiv:1507.02646.

\bibitem[\protect\citeauthoryear{Vehtari, Gelman, and Gabry}{Vehtari
  et~al.}{2017c}]{vehtari2017practical}
Vehtari, A., A.~Gelman, and J.~Gabry (2017c).
\newblock Practical {Bayesian} model evaluation using leave-one-out
  cross-validation and {WAIC}.
\newblock {\em Statistics and Computing\/}~{\em 27\/}(5), 1413--1432.
\newblock arXiv:1507.04544.

\bibitem[\protect\citeauthoryear{Wickham}{Wickham}{2009}]{ggplot2Rpackage}
Wickham, H. (2009).
\newblock {\em ggplot2: {E}legant Graphics for Data Analysis}.
\newblock Springer-Verlag New York.

\bibitem[\protect\citeauthoryear{Yao, Vehtari, Simpson, and Gelman}{Yao
  et~al.}{2018}]{yao2018yes}
Yao, Y., A.~Vehtari, D.~Simpson, and A.~Gelman (2018).
\newblock Yes, but did it work?: Evaluating variational inference.
\newblock {\em arXiv preprint arXiv:1802.02538\/}.

\end{thebibliography}

\clearpage
\section*{Supplementary Material: The 8-schools problem and the visualization of divergent trajectories}

\begin{figure}
\centering
\begin{subfigure}{0.48\textwidth}
\includegraphics[width=\textwidth]{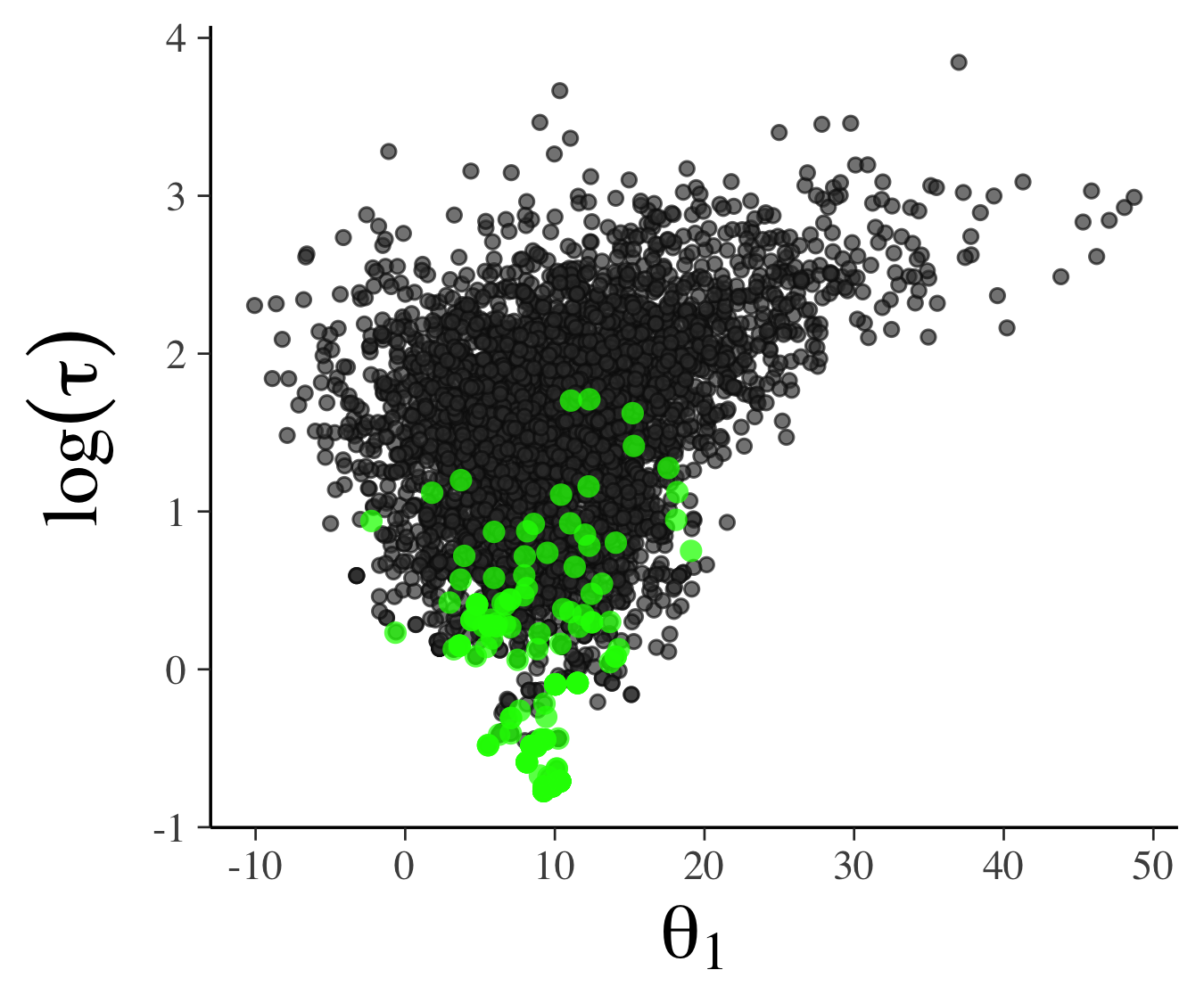}
\caption{A bivariate plot of the log standard deviation of school-level
effects ($\log{(\tau)}$, $y$-axis) against the mean for the first school
($\theta_1$, $x$-axis) for the 8-schools problem. The green dots indicate
starting points of divergent transitions.  The pile up of divergences
in a corner of the samples (in this case the neck of the funnel shape)
strongly indicates that there is a problem with this part of the parameter space.
This plot can be made using {\tt mcmc\_scatter} in \bayesplot.
}
\label{fig:mcmc_scatter_divs_schools}
\end{subfigure}
~
\begin{subfigure}{0.48\textwidth}
\includegraphics[width=\textwidth]{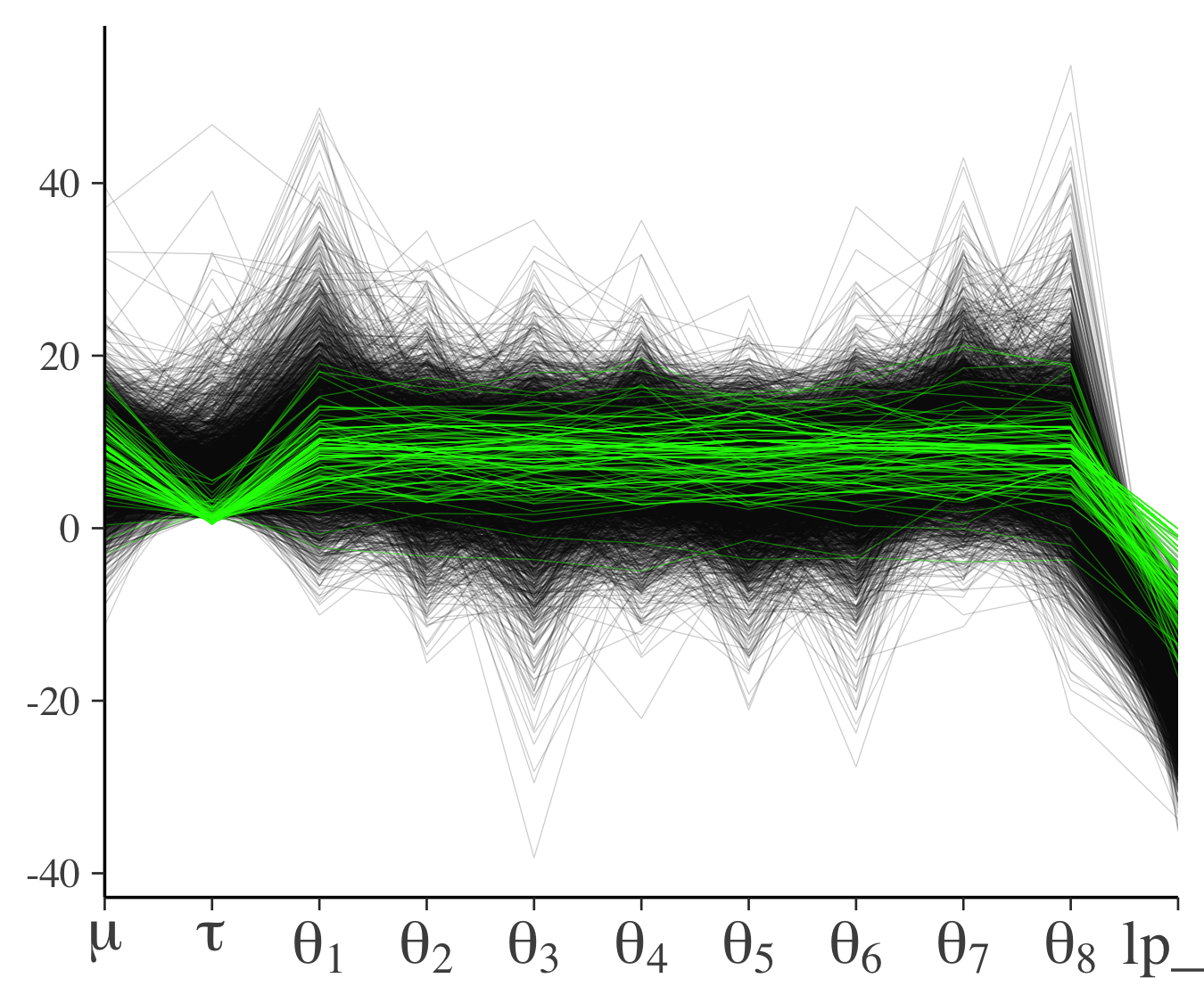}
\caption{Parallel coordinates plot for the 8-schools problem showing the
school-specific parameters ($\theta_1, \ldots, \theta_8$) and their
prior mean and standard deviation $(\mu, \tau)$.
The green lines indicate the starting points of divergent transitions. 
In this case it is clear that all of the divergent paths have a small value 
of $\tau$, which results in little variability in the $\theta_j$'s (the green 
lines are flat). This plot can be made using {\tt mcmc\_parcoord} in \bayesplot.}
\label{fig:mcmc_parcoord_divs_schools}
\end{subfigure}

\caption{\it Several different diagnostic plots for Hamiltonian Monte Carlo.
Models were fit using the RStan interface to Stan 2.17 \citep{rstan}.}
\end{figure}

Consider the hierarchical 8-schools problem outlined in \citep{rubin1981, gelman2013bda}. 
Figure \ref{fig:mcmc_scatter_divs_schools} shows a scatterplot of the log standard deviation 
of the school-specific parameters ($\tau$, $y$-axis) against the parameter representing the 
mean for the first school ($\theta_1$, $x$-axis). The starting points of divergent transitions, 
shown in green, concentrate in a particular region which is evidence of a geometric 
pathology in parameter space. 

Figure~\ref{fig:mcmc_parcoord_divs_schools}
gives a different perspective on the divergences. It is a parallel coordinates
plot including \emph{all} parameters from the 8-schools example with divergent
iterations also highlighted in green. We can see in both the bivariate plot and
the parallel coordinates plot that the divergences tend to occur when the
hierarchical standard deviation $\tau$ goes to 0 and the values of the
$\theta_j$'s are nearly constant. 

Now that we know precisely what part of the parameter space is causing problems,
we can fix it with a reparameterization \citep{betancourt2015}. Funnels in the parameter 
space can be resolved through a reparameterization that fattens out the problem area. 
The standard tool for fixing funnels caused by hierarchical models is moving to a non-centered
parameterization, where the narrowest coordinate is made a priori independent of
the other coordinates in the funnel \citep{betancourt2015}. This will typically
fatten out the funnel and remove the cluster of divergences.

\end{document}